\documentclass[%
 preprint,
 superscriptaddress,
 twocolumn,
 10pt,
 amsmath,amssymb,
 aps,
]{revtex4-2}

\usepackage{graphicx}% Include figure files
\usepackage{dcolumn}% Align table columns on decimal point
\usepackage{bm}% bold math
\usepackage{color}
\usepackage[hypertexnames=false]{hyperref} % add hypertext capabilities
\hypersetup{colorlinks,allcolors=black}
\usepackage{braket}
\usepackage[utf8]{inputenc}
\usepackage{microtype}

\newcommand{\vtg}{V_\mathrm{{TG}}}
\newcommand{\vbg}{V_\mathrm{{BG}}}
\newcommand{\bfA}{{\textbf{a}}}
\newcommand{\bfB}{{\textbf{b}}}
\newcommand{\bfC}{{\textbf{c}}}
\newcommand{\bfD}{{\textbf{d}}}
\newcommand{\bfE}{{\textbf{e}}}
\newcommand{\bfF}{{\textbf{f}}}

\begin{document}

%\preprint{APS/123-QED}

\title{Tunable quantum confinement of neutral excitons using electric fields and exciton-charge interactions}

\author{Deepankur Thureja}
\affiliation{%
 Institute for Quantum Electronics, ETH Zurich, Zurich, Switzerland\\
}%
\affiliation{
 Optical Materials Engineering Laboratory, Department of Mechanical and Process Engineering, ETH Zurich, Zurich, Switzerland
}%
\author{Atac Imamoglu}%
\email{imamoglu@phys.ethz.ch}
\affiliation{%
 Institute for Quantum Electronics, ETH Zurich, Zurich, Switzerland\\
}%
\author{Tomasz Smolenski}%
\affiliation{%
 Institute for Quantum Electronics, ETH Zurich, Zurich, Switzerland\\
}%
\author{Alexander Popert}%
\affiliation{%
 Institute for Quantum Electronics, ETH Zurich, Zurich, Switzerland\\
}%
\author{Thibault Chervy}%
\affiliation{%
Institute for Quantum Electronics, ETH Zurich, Zurich, Switzerland\\
}%
\author{Xiaobo Lu}%
\affiliation{%
Institute for Quantum Electronics, ETH Zurich, Zurich, Switzerland\\
}%
\author{Song Liu}%
\affiliation{%
Department of Mechanical Engineering, Columbia University, New York, NY, USA\\
}%
\author{Katayun Barmak}%
\affiliation{%
Department of Applied Physics and Applied Mathematics, Columbia University, New York, NY, USA\\
}%

\author{Kenji Watanabe}%
\affiliation{
 National Institute for Materials Science, Tsukuba, Ibaraki, Japan
}%
\author{Takashi Taniguchi}%
\affiliation{
 National Institute for Materials Science, Tsukuba, Ibaraki, Japan
}%
\author{David J. Norris}%
\affiliation{
 Optical Materials Engineering Laboratory, Department of Mechanical and Process Engineering, ETH Zurich, Zurich, Switzerland
}%
\author{Martin Kroner}%
\affiliation{%
 Institute for Quantum Electronics, ETH Zurich, Zurich, Switzerland\\
}%
\author{Puneet A. Murthy}%
\email{murthyp@phys.ethz.ch}
\affiliation{%
 Institute for Quantum Electronics, ETH Zurich, Zurich, Switzerland\\
}%

\maketitle

\textbf{Quantum confinement is the discretization of energy when motion of particles is restricted to length scales smaller than their de Broglie wavelength. The experimental realization of this effect has had wide-ranging impact in diverse fields of physics and facilitated the development of new technologies. 
In semiconductor physics, quantum confinement of optically excited quasiparticles, such as excitons or trions, is typically achieved by modulation of material properties \cite{Davies1997} – an approach crucially limited by the lack of in-situ tunability and scalability of confining potentials. Achieving fully tunable quantum confinement of optical excitations has therefore been an outstanding goal in quantum photonics.
Here, we demonstrate electrically controlled quantum confinement of neutral excitons in a gate-defined monolayer p-i-n diode. A combination of dc Stark shift induced by large in-plane fields and a previously unknown confining mechanism based on repulsive interaction between excitons and free charges ensures tight exciton confinement in the narrow neutral region. Quantization of exciton  motion manifests in multiple discrete, spectrally narrow, voltage-dependent optical resonances that emerge below the free exciton resonance. Our measurements reveal several unique physical features of these quantum confined excitons, including an in-plane dipolar character, one-dimensional center-of-mass confinement, and strikingly enhanced exciton size in the presence of magnetic fields.  
Our method provides an experimental route towards creating scalable arrays of identical single photon sources, which will constitute building blocks of strongly correlated photonic systems.}

\begin{figure*}[ht!]
	\includegraphics[width=12.5cm]{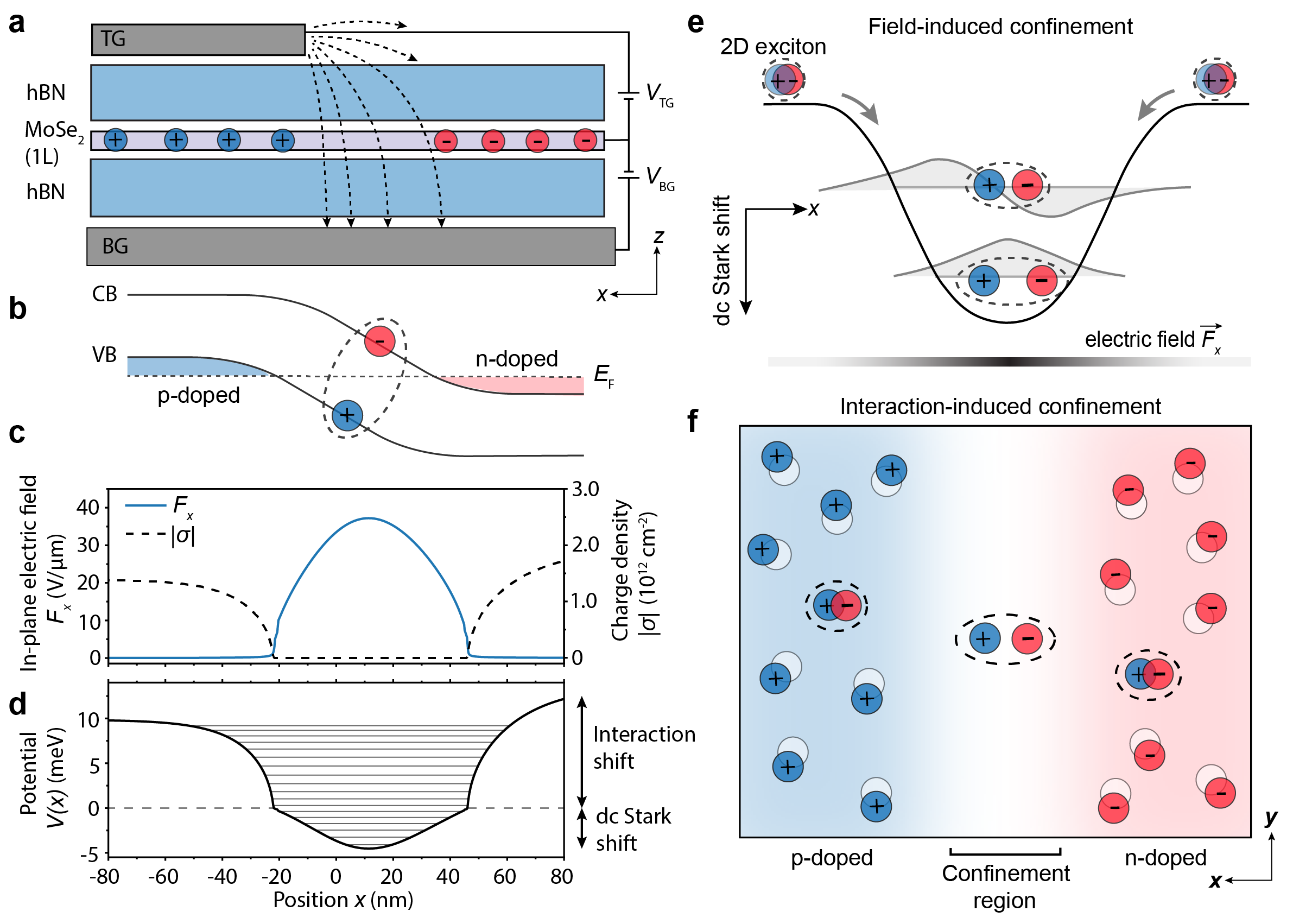}
	\caption{\textbf{Tunable quantum confinement of excitons in a lateral p-i-n junction.} (\textbf{a}) Schematic side view of the device consisting of a monolayer MoSe$_2$ encapsulated by h-BN and partially overlapping gate electrodes (TG and BG). By applying appropriate voltages between TG and BG, we generate inhomogeneous in-plane electric fields along the TG edge (black dashed lines), and obtain a p-i-n charging configuration, where holes and electrons are represented by blue and red circles, respectively. (\bfB) Schematic diagram of conduction and valence band edges in the p-i-n regime. (\textbf{c}) The spatial dependence of the in-plane electric field strength $F_x$ (blue line) and the total charge density $\sigma$ (dashed black line) obtained from electrostatic simulations. Exciton confinement stems from two fundamentally distinct effects, namely field-induced confinement and interaction-induced confinement, which are illustrated in (\textbf{e}) and (\textbf{f}) respectively. (\bfE) Field-induced confinement originates from the local in-plane electric field $F_x$, which induces a permanent dipole moment for 2D excitons and causes a dc Stark shift $\Delta E \propto |F_x|^2$. Therefore, the lower lying confined levels exhibit a larger dipole size. (\textbf{f}) Interaction-induced confinement arises from the repulsive interaction between excitons and itinerant charges, which imposes a density-dependent energy cost, $\Delta E \propto \sigma(x)$ for excitons entering the doped regions. This results in a repulsive potential barrier that pushes the excitons back towards the neutral region. The combined simulated potential including dc Stark shift and interaction shift is shown in (\textbf{d}), along with the discrete eigenstates of excitons. 
	} 
	\label{fig:concept}
\end{figure*}

Neutral excitons, which are bound electron-hole pairs,  are intrinsically more challenging to electrically confine than charged particles. One possible route for exciton confinement involves the dc Stark effect which ensures that excitons see an attractive potential around the absolute  maximum of an inhomogeneous electric field distribution. However, achieving quantum confinement in such a potential requires that the energy splitting between discrete motional excitonic states ($\hbar\omega$) exceed the exciton line broadening $\Gamma$ as well as the characteristic energy of thermal fluctuations $k_{\mathrm{B}}T$. For excitons in semiconductor heterostructures at $T=4$~K, this implies $\hbar\omega \gtrsim 1\,$meV. This in turn requires a confinement length scale of  $\ell = \sqrt{\hbar/m_X\omega} \lesssim 10\,$nm for exciton mass $m_X$ comparable to free electron mass. Engineering electrically tunable confinement potentials at such small length scales is a technical challenge. In addition, unless the exciton binding energy is much larger than $\hbar\omega$, the requisite applied fields will lead to fast ionization of excitons, drastically reducing their radiative efficiency. 

Previous experiments have mainly approached the problem of electrical confinement by relying on indirect excitons, where the electron and hole comprising an exciton are spatially separated in different quantum wells. This gives rise to a permanent electric dipole moment that couples more strongly to the applied fields. In III-V semiconductor heterostructures with coupled quantum wells, different potential landscapes have been demonstrated for indirect excitons, such as ramps, lattices, and harmonic traps \cite{Hagn1995,Rapaport2005,Gartner2007,Schinner2013,Butov2017}. However, the requirement to suppress exciton ionization in these systems \cite{Hammack2006} prevented the observation of quantum confinement. Moreover, the quantum wells typically need to be buried deep within the heterostructure, which limits the electric field gradients that can be applied using lithographically patterned gates outside the structure. Electrical manipulation of spatially indirect excitons at large length scales has also been reported in transition metal dichalcogenide (TMD) heterostructure devices \cite{Unuchek2018,Wang2018a,Liu2020,Jauregui2019}. Nevertheless, nanoscopic electrically defined potentials required to quantum confine excitons have so far not been realized in any experimental platform.  

In this work, we overcome these challenges and demonstrate an electrically tunable quantum confining potential for spatially direct excitons in a monolayer semiconductor. In achieving this goal, we not only solve a long-standing technical problem in quantum photonics, but reveal intriguing aspects of quantum confined excitons unique to our system. In addition to being highly relevant for technological applications, our work highlights that the electrically controlled quantum confined system is a promising experimental platform for exploring a rich variety of phenomena, such as interacting dipolar excitons and low dimensional excitonic gases.

\subsection*{Mechanisms of quantum confinement}
Our experimental scheme to confine excitons is illustrated in Fig.\,\ref{fig:concept}. We base our experiments on monolayer TMD heterostructures, as they offer several key advantages over conventional semiconductor platforms, the most important one being the ultra-strong exciton binding energy, which render the excitons resilient to large in-plane electric fields. We consider a device which includes a monolayer TMD semiconductor, such as MoSe$_2$, encapsulated by insulating dielectric spacer layers and two gate electrodes (top gate, TG; bottom gate, BG). Importantly the top and bottom gates have partial spatial overlap as shown in Fig.\,\ref{fig:concept}\,\bfA, which allows to separately define adjacent p- and n-doped regions, and generate in-plane electric fields along the TG edge. The energy of conduction (CB) and valence band (VB) edges relative to the Fermi level $E_F$ in this doping configuration are illustrated in Fig.\,\ref{fig:concept}\,\bfB. In Fig.\,\ref{fig:concept}\,{\bfC}, we show the spatial dependence of in-plane electric field $F_x$ (blue line) and total charge density $|\sigma(x)|$ (black dashed line) in the vicinity of the TG edge, obtained from electrostatic simulations of the device.

%Electrostatic simulations of such a p-i-n device provide a quantitative picture of the spatial and voltage dependence of the charge density ($\sigma(x)$) and in-plane electric field ($F_x$) distribution in the vicinity of the TG, as shown in Fig.\,\ref{fig:concept}\,{\bfC}. The simulations indicate that a neutral junction region as narrow as $20 - 30\,$nm and peak in-plane electric fields of $\sim 0.1\,\mathrm{V/nm}$ can be achieved in realistic devices with highly inhomogeneous in-plane electric fields and steep charge density gradients. A detailed description of the simulations can be found in \cite{som}.

Two fundamentally distinct effects contribute to quantum confinement of excitons in our structure. The first arises from the strong in-plane electric field $F_x$, which polarizes the excitons along $x$ and lowers their energy due to a dc Stark shift $\Delta E_S = -\frac{1}{2} \alpha |F_x|^2$, where $\alpha$ is the exciton polarizability \cite{Cavalcante2018}. Since $F_x$ vanishes on either side of the i-region (Fig.\,\ref{fig:concept}\,\bfC), excitons experience an attractive confining potential towards the local maximum in $|F_x(x)|$. Hence, we expect the dipole size to be the largest for excitons confined in the lowest eigenstates (see Fig.\,\ref{fig:concept}\,\bfE). In addition to the Stark shift, we describe a new confinement mechanism that stems from the interaction between neutral excitons and itinerant charges present in the neighbouring p- and n-doped regions (Fig.\,\ref{fig:concept}\,\bfF). As neutral excitons generated in the i-region enter the n- or p- doped regions, they experience a repulsive interaction which increases their energy proportional to the charge density, $\Delta E_P \propto \sigma(x)$. This many-body dressed state can be described as a repulsive polaron \cite{Efimkin2017}, and has been previously observed in charge-tunable semiconductor heterostructure devices \cite{Sidler2017}. In this scenario, a gradient in charge density exerts a force on excitons \cite{Chervy2020} that pushes them towards the local minimum in charge density. A steep charge density gradient on both sides of the i-region therefore acts as a repulsive potential barrier for excitons, which confines them in the i-region. Remarkably, this effect leads to confinement even for exciton energies larger than the free, zero-momentum 2D exciton energy ($E_\mathrm{X,2D}$).

The total potential experienced by excitons in the center-of-mass (COM) frame is a sum of dc Stark shift and repulsive interaction shift contributions:
\begin{equation}
    V(x) = \underbrace{-\frac{1}{2}\alpha |F_x(x)|^2}_\text{dc Stark shift} \,\,\,\, + \underbrace{\beta |\sigma(x)|}_\text{interaction shift},
    \label{eqn:potential}
\end{equation}
Here, the proportionality constant $\beta$ is an effective exciton-charge coupling constant. We emphasize that the potential in Eq.\,\ref{eqn:potential} provides confinement only for excitons whereas unbound electrons or holes experience a repulsive potential that accelerates them towards the n- and p-doped regions, respectively. The relative contribution of the field-induced and interaction-induced confinement mechanisms can be controlled by tuning the charge densities in the neighbouring regions. In addition, even the asymmetry of the potential can be controlled by independent tuning of the electron and hole densities using the two gates. The exciton confinement strength achieved with the potential in Eq.\,\ref{eqn:potential} depends on the geometry of the device and excitonic properties in the material. We calculate the potential using the following parameters in the electrostatic simulations: $\alpha = 6.5\,\mathrm{eV\,nm}^2/\mathrm{V}^2$ (for MoSe$_2$) \cite{Cavalcante2018}, $m_X = 1.3m_e$ \cite{Larentis2018,Zhang2014,Goryca2019}, and h-BN thickness (top and bottom) of $30\,$nm. We use exciton-electron coupling strength $\beta = 0.7\,\mu\mathrm{eV}\mu\mathrm{m^2}$, which is experimentally determined from the density-dependence of the repulsive polaron energy \cite{som}.  In Fig.\,\ref{fig:concept}\,\bfD, we show the overall potential for bottom gate voltage $\vbg = 4\,$V and top gate voltage $\vtg = -8\,$V. The contributions from field- and interaction-induced confinement mechanisms are indicated. We thus  obtain a level separation between the discrete eigenstates $\hbar\omega_x \approx 1\,$meV ($\ell \approx 7.5\,$nm), which is on the order of the 2D  exciton radiative linewidth $\Gamma$. The results of simulations for different voltages are shown in Extended Fig.\,\ref{fig:Fx_nc}.

\begin{figure*}[ht!]
	\includegraphics[width=11.5cm]{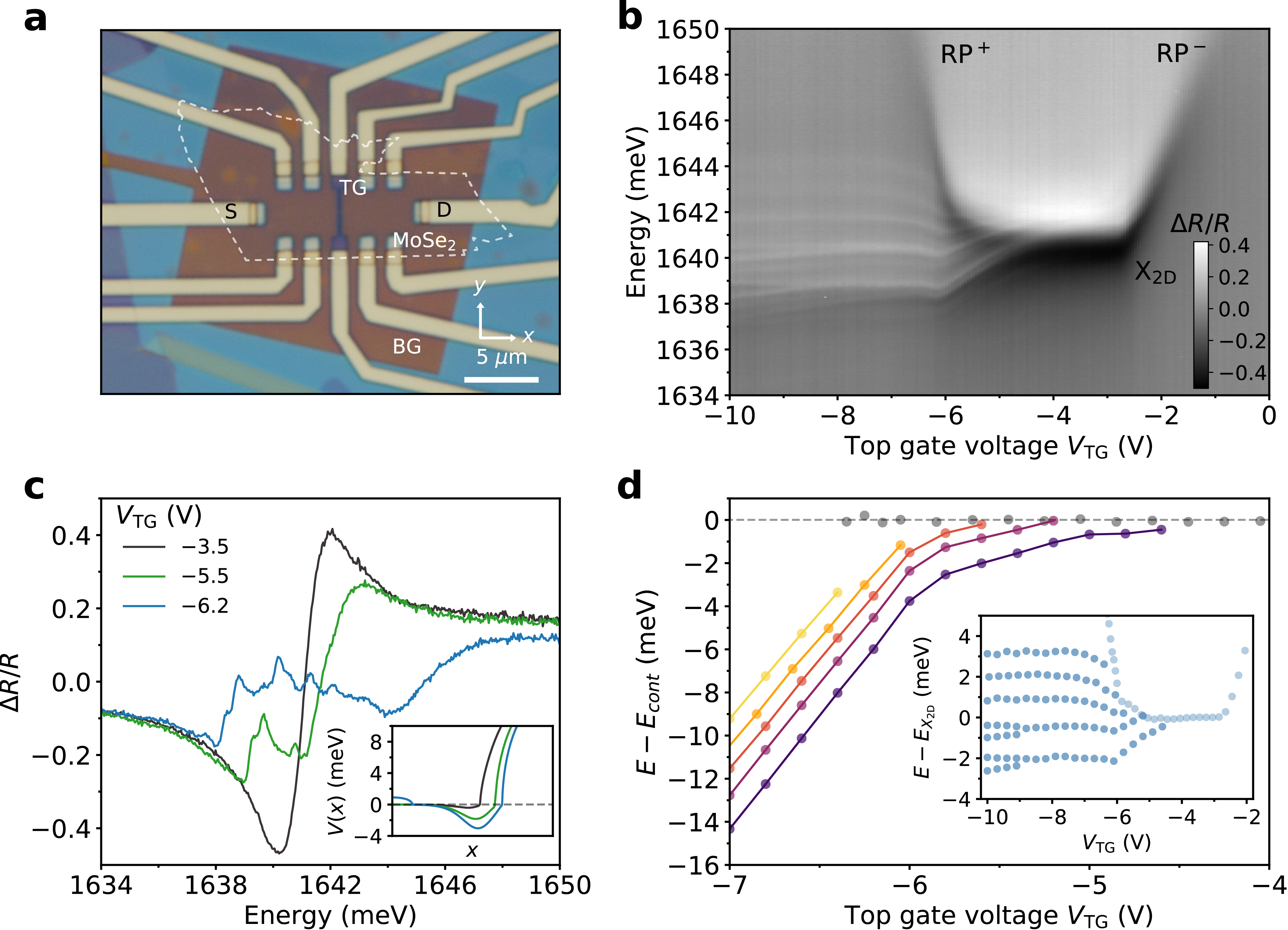}
 	\caption{\textbf{Optical signatures of quantum confined excitons.}  (\bfA) Optical micrograph of Device 1, where the dashed white line indicates the monolayer MoSe$_2$ flake. S and D refer to source and drain electrodes in the device. (\bfB) The normalized differential reflectance $\Delta R/R$ as a function of $\vtg$, taken at the TG edge at $\vbg = 4\,$V. In addition to the typical features associated with the neutral exciton (X$_\mathrm{2D}$) and repulsive polaron branches (RP$^+$ and RP$^-$) from underneath the TG, we observe narrow discrete spectral lines that red shift as a function of $\vtg$ compared to the exciton or RP$^+$ energy at that voltage. (\bfC) Spectra at $\vtg = -3.5\,$V (black), $-5.5\,$V (green), and $-6.2\,$V (blue). The inset shows the estimated shape of the total potential $V(x)$ for excitons at the corresponding voltages (see Eq.(\ref{eqn:potential})). (\bfD) Energy of discrete resonances ($E$) with respect to the free exciton continuum $E_{\mathrm{cont}}$. The continuum energy is the 2D exciton energy in the neutral regime, and the repulsive polaron energy in the doped regime. The inset shows the resonance energies obtained from Lorentzian fits of reflectance with respect to the 2D exciton energy. We observe a maximum level spacing of $\hbar\omega \sim 1.5\,$meV between the lowest confined states, accompanied by a narrowing of the linewidth from $\Gamma_\mathrm{2D} \approx 2\,$meV for free excitons, to $\Gamma \approx 300\,\mu$eV for confined excitons. } 
	\label{fig:WL}
\end{figure*} 

\subsection*{Observation of exciton quantum confinement}

We fabricate two vdW heterostructure devices implementing the concept elaborated above. The first device (Device 1) consists of a monolayer MoSe$_2$ encapsulated by $30\,$nm thick h-BN flakes and contacted with palladium electrodes. This device features a gold BG encompassing the entire MoSe$_2$ flake that allows for global doping of the semiconductor, and a $13\,$nm-thick and $200\,$nm wide gold TG running across the flake which allows to locally modify the charge density underneath. The thin TG is optically transparent, allowing to probe the local optical properties of the region underneath the TG and along its edge. An optical micrograph of Device 1 is shown in Fig.\,\ref{fig:WL}\,\bfA, where the outline of the MoSe$_2$ monolayer is marked with a dashed white line. Our second device (Device 2) is also a dual-gated heterostructure similar to Device 1, except fabricated with few-layer graphene gates instead of gold and having different h-BN thicknesses (top h-BN: $40\,$nm; bottom h-BN: $54\,$nm). The characterization of both devices is presented in Extended Figs.\,\ref{fig:Dev1} and \ref{fig:Dev2}. Unless otherwise stated, we will focus on experimental data from Device 1. All measurements are performed at liquid helium temperature. 

The modification of excitonic states due to confinement is revealed in the optical response of the narrow depleted region around the TG. We perform optical reflectance spectroscopy by positioning the optical spot on the TG edge \cite{som}. Due to the diffraction-limited spot size of our optical setup, our measurements correspond to the combined optical response of three distinct spatial regions: (I) the electron-doped region away from the TG that is affected only by the BG, (II) the region directly underneath the TG, and (III) the narrow region between I and II. The contribution of region I to the total optical response remains unchanged as $\vtg$ is varied. Therefore, to discern the influence of the TG alone, we measure $\vtg$-dependent spectra for fixed values of $\vbg$, and subtract the reflectance spectrum obtained for $\vtg = 0$\,V from the total signal. This results in the normalized differential reflectance, 

\begin{equation}
\frac{\Delta R}{R} = \frac{R(V_{\mathrm{TG}}) - R(V_{\mathrm{TG}} = 0)}{R(V_{\mathrm{TG}} = 0)}.
\label{eqn:drr}
\end{equation}

In Fig.\,\ref{fig:WL}\,\bfB, we present ${\Delta R}/{R}$ as a function of $\vtg$ at fixed $\vbg = 4\,$V, which corresponds to an electron density $\sigma_n = 2 \times 10^{12}\,\mathrm{cm}^{-2}$ in region I. First, we identify the typical doping-dependent optical response from region II directly underneath the TG. This includes a neutral regime ($X_\mathrm{2D}$: $-6\,\mathrm{V} \lesssim \vtg \lesssim -3\,\mathrm{V}$) flanked by repulsive polaron branches on the electron (RP$^-$: $\vtg \gtrsim -3\,$V) and hole (RP$^+$: $\vtg \lesssim -6\,\mathrm{V}$) doped sides, which blue shift with respect to the neutral exciton state. Interestingly, the fact that we observe a hole-side repulsive polaron branch RP$^+$ shows that, hole doping of region II is possible even without direct electrical contacts. This is an optical doping effect that stems from the dissociation of excitons due to strong in-plane fields in the vicinity of the TG edge \cite{som}. As the excitons are dissociated with a finite probability, the free holes and electrons are accelerated towards region II and I, respectively.

\begin{figure*}[ht!]
	\includegraphics[width=12.2cm]{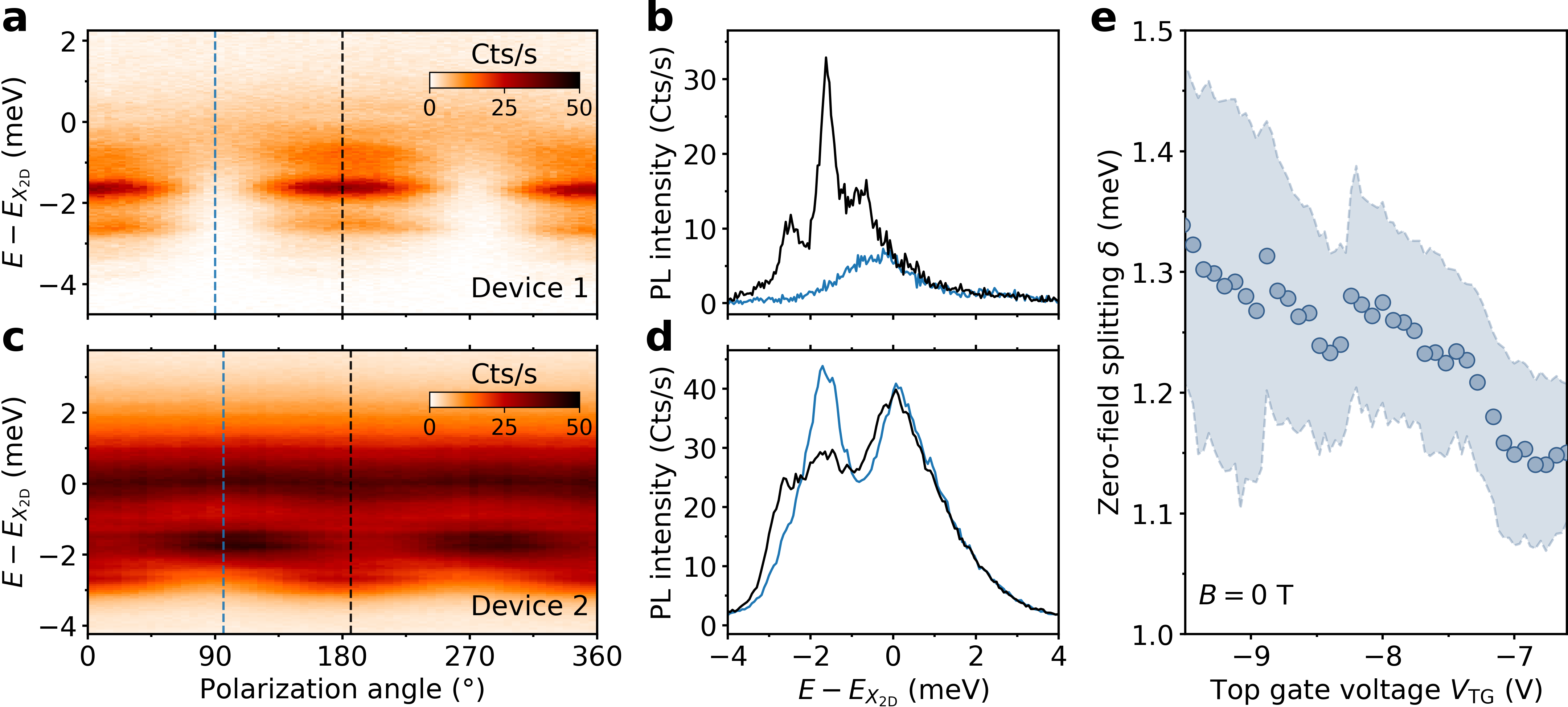}
	\caption{\textbf{Polarization dependence and 1D confinement.} (\bfA, \bfC) PL spectra as a function of linear polarization angle in Device 1 and Device 2, respectively. (\bfB, \bfD) Spectra at orthogonal polarization angles for Device 1 and 2, respectively. In Device 1, all discrete states show linear polarized states along the edge of the gate, with a high degree of linear polarization. In Device 2, the confinement is weaker due to thicker h-BN spacer layers, which leads to the observation of both $x-$ and $y-$ polarized states with a finite energy splitting $\delta \sim 1\,$meV. (\bfE) The polarization splitting $\delta$ increases with decreasing $\vtg$, corresponding to tighter confinement. The shaded blue area represents fitting errors.
	}  
	\label{fig:pol}
\end{figure*}

In addition to the expected optical response, we observe multiple narrow and discrete spectral lines for $\vtg \lesssim -4$\,V, which emerge from the 2D exciton and repulsive polaron continuum, and red shift with decreasing $\vtg$. We confirm that the narrow lines only exist along the edge of the TG by performing position-dependent reflectance and photoluminescence (PL) measurements on both Device 1 and 2 (see Extended Fig.\,\ref{fig:1Dedge}). In Fig.\,\ref{fig:WL}\,\bfC, we show representative reflectance spectra taken at $\vtg = -3.5\,$V, $-5.5\,$V and $-6.2\,$V which highlight the appearance of new states with varying $\vtg$. For reference, estimates of the corresponding potential $V(x)$ (Eq.\,(\ref{eqn:potential})) for the respective voltages are shown in the inset. The resonance center frequencies of the discrete states, obtained from Lorentzian fits to reflectance spectra are shown in the inset of Fig.\,\ref{fig:WL}\,\bfD. We extract an energy separation $E_2 - E_1 \sim 1.5\,$meV between the lowest two resonances at $\vtg = -8\,$V. 
% We observe these discrete signatures also in PL measurements as shown in Extended Fig.\,\ref{fig:PLWL}.

The emergence of discrete lines from the 2D continuum can be attributed to the quantization of the COM motion of excitons due to strong confinement. The level separations observed in the experiment are similar to those obtained from the electrostatic simulations of the device \cite{som}. These observations are strongly corroborated by optical measurements on Device 2 (see Extended Fig.\,\ref{fig:Dev2}\,\bfC). Further evidence for strong confinement of excitons is provided by the significant reduction of the linewidth of the discrete states compared to the free exciton. Whereas the 2D exciton exhibits a linewidth of $\Gamma \sim 2\,$meV, the lowest discrete resonance has a linewidth $\Gamma \sim 300\,\mu$eV. Such narrowing is qualitatively expected to stem from three factors:  (i) lower inhomogeneous broadening since exciton COM motion is restricted to a smaller spatial area due to confinement; (ii) reduction of radiative decay of confined excitons as compared to their free 2D counterparts according to the ratio $\ell_x/\lambda_\mathrm{photon}$, where $\ell_x$ is the harmonic oscillator length along $x$ and $\lambda_\mathrm{photon}$ is the photon wavelength; (iii) the reduced electron-hole wavefunction overlap originating from the permanent in-plane electric dipole moment induced by the in-plane electric field, further reducing their radiative decay rate. In addition, we emphasize that the reduced linewidth of the confined exciton resonances shows that the non-radiative decay due to field-induced ionization remains negligible in the explored range of parameters. 

%Furthermore, the reduced linewidth of the confined exciton resonances demonstrates their resilience towards ionization.

In the voltage regime $-6\,\mathrm{V} < \vtg < -4\,\mathrm{V}$, which corresponds to the i-i-n charging configuration, the red shift of the narrow resonances with decreasing $\vtg$ arises mainly from the field-induced confinement (Fig.\,\ref{fig:concept}\,\bfE) mechanism, wherein the dc Stark shift lowers the energy below the continuum given by the 2D exciton energy $E_\mathrm{X,2D}$. In the p-i-n regime ($\vtg < -6\,$V), as region II becomes hole doped, we observe additional quantized modes with energy higher than the $E_\mathrm{X,2D}$, that split off from the repulsive-polaron branch RP$^+$ (inset of  Fig.\,\ref{fig:WL}\,\bfD). In this regime, the confinement potential is the sum of the field-induced and interaction-induced contributions. Therefore, the free-particle continuum is no longer the 2D exciton state but the blue shifted repulsive polaron (RP$^+$) in region II. In other words, to escape the confinement, an exciton in the lowest state must pay not only the dc Stark energy shift, but an additional repulsive polaron energy, $E = \beta \cdot \sigma_\mathrm{max}$, where $\sigma_\mathrm{max}$ denotes the maximum charge density. The energy of confined resonances with respect to the continuum energy ($E_\mathrm{cont}$) at each $\vtg$, as shown in Fig.\,\ref{fig:WL}\,\bfD, allows to observe the successive emergence of discrete states below the free particle continuum as the potential is made deeper. 

\begin{figure*}[ht!]
	\includegraphics[width=14cm]{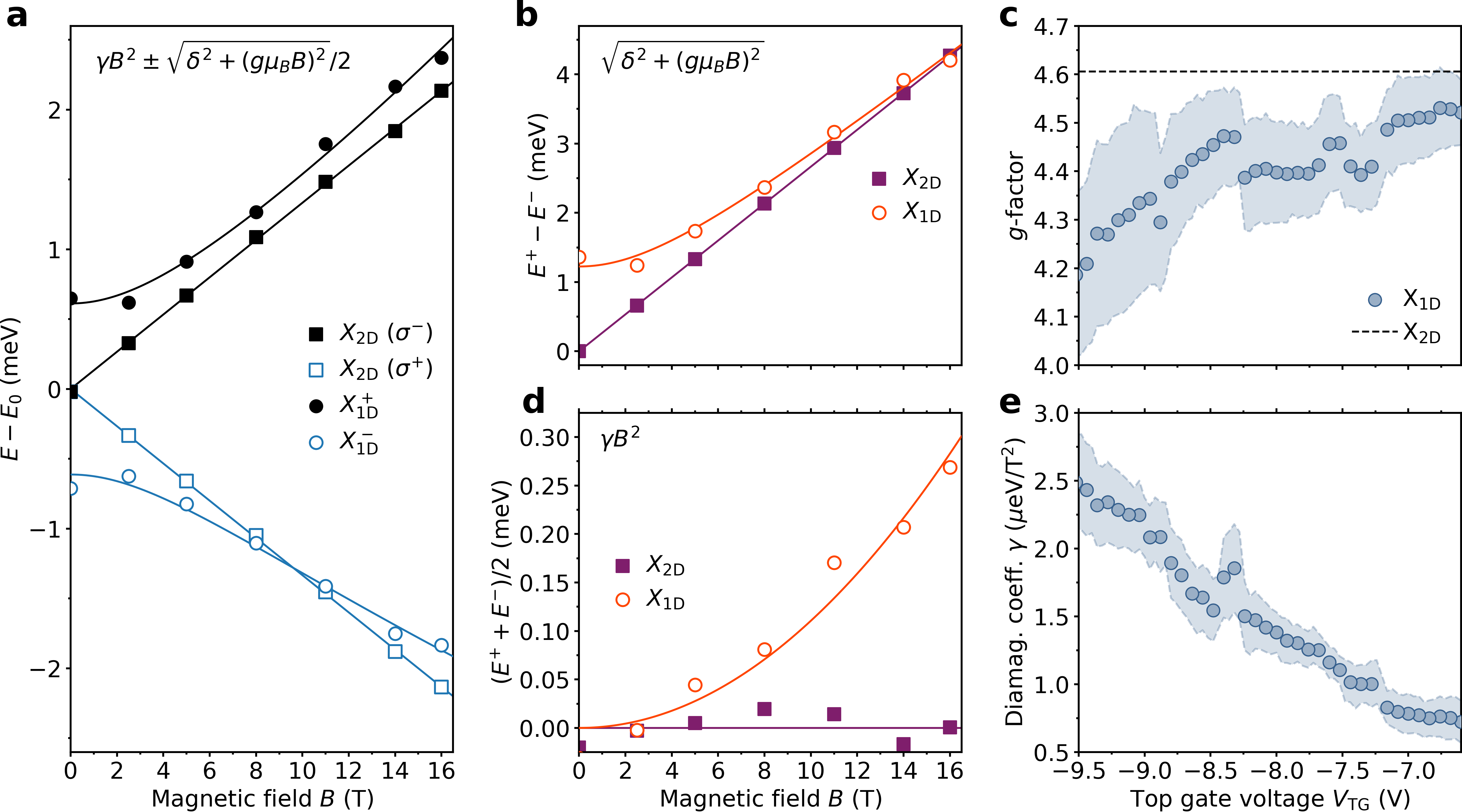}
	\caption{\textbf{Internal structure of confined excitons.} (\bfA) Polarization splitting as a function of magnetic field for 2D excitons ($X_\mathrm{2D}$, black closed and blue open squares) and 1D confined excitons ($X_\mathrm{1D}$, black closed and blue open circles) in Device 2. Here, $\gamma$ is the diamagnetic coefficient, $\delta$ is the zero-field splitting and $g$ is the exciton g-factor. These measurements are all taken with Device 2 at $\vbg = 1\,$V. The B-field dependent $\sigma_+$ and $\sigma_-$-polarized states can be fit with the function $E_{\pm} = \gamma B^2 \pm \sqrt{\delta^2 + (g\mu_EB^2)}/2$. (\bfB) The energy splitting $E^+ - E^-$ shows linear behavior at large fields, which results in a g-factor $g = 4.6$ for $X_{2D}$ and voltage dependent $g$ for $X_{1D}$ that reduces with decreasing $\vtg$, as shown in (\bfC). (\bfD) The average energy $(E^+ + E^-)/2$ represents the diamagnetic shift of excitons. The 2D exciton shows no measurable diamagnetic shift, even at 16 T (purple squares), consistent with the small Bohr radius. On the other hand, the 1D dipolar excitons exhibit a large quadratic shift with B-field confirming that electric field-induced confinement increases the exciton size. (\bfE) The diamagnetic coefficient for 1D excitons, obtained from quadratic fits to (\bfD), increases dramatically with decreasing $\vtg$. The shaded blue area in (\bfC) and (\bfE) represents fitting errors.
	}  
	\label{fig:Bfield}
\end{figure*} 

\subsection*{One-dimensional confining potential}
The strong in-plane confinement induced perpendicular to the TG edge implies that exciton confinement in our system is effectively one-dimensional in nature. To reveal the effect of such an anisotropic potential on the COM motion of excitons, we turn our attention to the polarization properties of emission from the confined states. Excitons in 1D semiconductor nanowires have been previously reported to emit photons that are linearly polarized along the wire axis \cite{Wang2001,Akiyama1996,Lefebvre2004,Bai2020}. 
%It is well known that excitons in 1D semiconductor nanowires emit photons that are linearly polarized along the wire  \cite{Wang2001,Akiyama1996,Lefebvre2004,Bai2020}. 
Linearly polarized emission originates from long-range electron-hole exchange interaction, which couples the valley degrees of freedom and the COM motion of excitons \cite{Glazov2015,Yu2015}. For finite exciton COM momenta, the valley-COM coupling leads to an energy splitting between longitudinal and transverse electromagnetic modes. Introducing spatial anisotropy, for instance in the form of a 1D confinement potential, breaks the rotational symmetry of the system, which in turn opens a gap between orthogonal linear polarization states at $k=0$. The magnitude of the polarization splitting $\delta$ depends on the COM momentum-space wave function $\psi(k)$ and the electron-hole exchange coupling strength $J$ \cite{Bai2020}, according to $\delta = (J/K)\int \psi^2(k)|k|dk$, where $K$ is momentum separation between $\Gamma$ and $K$ points of the Brillouin zone. In our system, stronger confinement (larger COM momentum spread) is accompanied by larger dipole size (smaller electron-hole overlap), and the polarization splitting is determined by the interplay of these two effects. 

%For typical experimental parameters: oscillator length $\ell_x \sim 7 - 30\,$nm and dipole size of $0.1 - 0.35\,$nm, we estimate a polarization splitting of the order of $1 - 4\,$meV.  

%The light-matter coupling $J$ depends on the electron-hole overlap in relative coordinates, hence a larger dipole moment leads to smaller $\delta$. At the same time, stronger confinement leads to larger COM momentum spread and hence larger $\delta$. 

%i.e. smaller oscillator length (tighter confinement) should lead to larger valley hybridization at large momenta and hence larger linear polarization splitting \cite{som}. 

We measure linear polarization-resolved photoluminescence (PL) spectra of quantum confined excitons in Device 1 and 2 (see Fig.\,\ref{fig:pol}). In Device 1 (panels {\bfA} and  {\bfB}), we find all discrete states to be linearly polarized parallel to the TG edge (i.e. polarized along $y$). Moreover, the emission from confined states exhibits a high degree of linear polarization $\xi = (I_\parallel - I_\perp)/(I_\parallel + I_\perp) \sim 0.8$, where $I_\parallel$ and $I_\perp$ are the intensities of polarization parallel and perpendicular to the wire. On the other hand, in Device 2 (panels {\bfC} and {\bfD}), we observe a polarization doublet comprising both parallel ($y-$polarized) and perpendicular ($x-$polarized) components with an energy splitting of $\delta = E_\parallel - E_\perp \sim 1\,$meV. The discrepancy between the polarization-resolved measurements of the two devices can be mostly explained by their different designs. Device 2 consists of substantially thicker h-BN spacer layers than Device 1 and hence a larger separation between top and bottom gates. Consequently, the exciton confinement is expected to be weaker in Device 2. While in Device 1, the confinement is strong enough that $x-y$ polarization splitting exceeds the confinement energy, the weaker confinement in Device 2 allows to observe both polarization states and their relative splitting $\delta$. Our explanation is further supported by the observation that as we make the confinement tighter in Device 2 by reducing $\vtg$, the polarization splitting $\delta$ increases, as shown in Fig.\,\ref{fig:pol}\,\bfE. These observations confirm our picture of excitons trapped in highly anisotropic 1D potentials.

\subsection*{Internal structure of quantum confined excitons}
The presence of strong inhomogeneous in-plane electric fields in the i-region influences not only the COM motion, but also the relative wavefunction of the exciton. As illustrated in Fig.\,\ref{fig:concept}\,\bfE, lowest energy excitons experience the strongest electric field at the trap center, and therefore should possess the largest dipole moment. To shed more light on the internal structure of quantum confined excitons, we perform circular polarization-resolved measurements with an external magnetic field $B$, applied perpendicular to the plane. The combination of in-plane electric and out-of-plane magnetic fields has unexpected consequences for the excitonic structure as we show below.
% In this scenario, the Hamiltonian for an interacting electron and hole in the relative frame of the exciton is given by:
% \begin{equation}
%     \hat{H} = \frac{\hbar^2}{2\m_r}\Delta_r + i\frac{e \hbar B}{}
% \end{equation}
In the perturbative regime, applying a $B-$field lifts the valley degeneracy and induces a splitting between circular polarization states ($\sigma^+$, $\sigma^-$), which leads to the following $B-$field dependence of the energies:
\begin{equation}
    E^{\pm} =  E_0 + \gamma B^2 \pm \sqrt{\delta^2 + (g\mu_\mathrm{B}B)^2}/2.
\end{equation}
Here, $E_0$ is the average exciton energy at $B=0\,\mathrm{T}$, $\delta$ is the zero $B-$field polarization splitting, $g$ is the exciton g-factor, $\mu_\mathrm{B}$ is the Bohr magneton, and $\gamma$ is the diamagnetic coefficient. In Fig.\,\ref{fig:Bfield}\,\bfA, we show the energies of $E^+$ and $E^-$ as a function of $B$-field for both 2D excitons ($X_\mathrm{2D}$) and 1D confined excitons ($X_\mathrm{1D}$) in Device 2. The energy splitting as a function of $B$-field is shown in Fig.\,\ref{fig:Bfield}\,\bfB. The 2D excitons show the expected behavior, i.e. perfect valley degeneracy ($\delta = 0$) at $B = 0\,$T, and linear energy splitting with increasing B-field. In contrast, the 1D excitons have a finite splitting $\delta \sim 1\,$meV at $B=0\,\mathrm{T}$ and approach linear dependence asymptotically only at high $B$-fields. For 1D excitons, the $g-$factor thus obtained decreases with decreasing $\vtg$ and is consistently lower than $g_\mathrm{2D} \sim 4.2$ for 2D excitons. We speculate that strong in-plane electric fields may modify the Bloch states of free electrons and holes, as well as the relative motion of bound electrons and holes in excitons, thus leading to the observed voltage-dependent shifts of the $g-$factor. 

Information on the spatial extent of the exciton is encoded in its diamagnetic properties. The rms size $\braket{r^2}$ of the 1s exciton is related to the diamagnetic energy shift according to $E_\mathrm{dia} = e^2 \braket{r^2} B^2/8m_\mathrm{r}$, where $m_\mathrm{r} = m_e m_h/(m_e + m_h)$ is the reduced mass. In Fig.\,\ref{fig:Bfield}\,\bfD\, we show the $B$-field dependence of the average energy of the two polarization states $E_\mathrm{dia} = (E^+ + E^-)/2$, which allows us to extract the diamagnetic coefficient $\gamma$ for both 2D (purple squares) and 1D confined excitons (orange circles). 2D excitons in TMD monolayers are strongly bound and have small Bohr radii ($a_\mathrm{B} \sim 1\,$nm for MoSe$_2$), which necessitates extremely large $B$-fields for diamagnetic shifts to be measurable \cite{Stier2016}. As a consequence, in our experiments performed at $B < 16\,$T we do not observe a sizeable shift of the average energy $(E^+ + E^-)/2$ for 2D excitons. On the other hand, we find that the 1D confined exciton states show unusually large diamagnetic shifts (orange circles) and corresponding diamagnetic coefficient extracted from quadratic fits. Furthermore, as shown in Fig.\,\ref{fig:Bfield}\,\bfE, $\gamma$ increases dramatically with decreasing $\vtg$ (or increasing in-plane electric fields), reaching values up to $2.5\,\mu\mathrm{eV}/\mathrm{T}^2$. Such values of $\gamma$ imply an anomalously large exciton size  $\sqrt{\braket{r^2}} \sim 6\,$nm, which is in turn comparable to the confinement length scale of the COM motion. To the best of our knowledge, such large diamagnetic shifts for ground-state neutral excitons in TMD heterostructures have not been previously reported.

%In Fig.\,\ref{fig:Bfield}\,\bfD\, we show the $B$-field dependence of the average energy of the two polarization states $E_\mathrm{dia} = (E^+ + E^-)/2$. This allows to extract the diamagnetic coefficient $\gamma$ of excitons for both 2D (purple squares) and 1D excitons (orange circles). The exciton diamagnetic shift is expected to grow quadratically with $B$. This allows to infer the spatial size of the 1s excitonic relative wavefunction, given by $E_\mathrm{dia} = e^2 \braket{r^2} B^2/8m_\mathrm{r}$, where $\braket{r^2}$ is the rms size of the $1s$ exciton and $m_\mathrm{r} = m_e m_h/(m_e + m_h)$ is the reduced mass. 2D excitons in TMD monolayers are strongly bound and have small Bohr radii ($a_\mathrm{B} \sim 1\,$nm), which necessitates extremely large $B$-fields for diamagnetic shifts to be observable. As a consequence, in our experiments performed at $B < 16\,$T we do not observe a sizeable shift of the 2D exciton. On the other hand, the 1D exciton states show strikingly large diamagnetic shifts which grow quadratically with $B$-field (solid orange curve), thus allowing to extract the diamagnetic coefficient $\gamma$. In Fig.\,\ref{fig:Bfield}\,\bfE, we show $\gamma$ as a function of $\vtg$. Remarkably, we observe that the diamagnetic coefficient $\gamma$ increases dramatically with larger in-plane electric fields reaching values upto $2.5\,\mu\mathrm{eV}/\mathrm{T}^2$. This implies an anomalously large exciton size of $\braket{r^2} \sim 8\,$nm.

Although a complete theoretical understanding of the large diamagnetic shift of confined excitons in our system is currently lacking, we provide one possible phenomenological explanation that may provide hints for future works. We note that in the presence of both in-plane electric and magnetic fields, the electron (or hole) in the relative frame of the exciton experiences a superposition of three potentials: (a) the Coulomb potential $V_C(r) = -e^2/4\pi\epsilon r$, (b) the diamagnetic potential $V_\mathrm{dia}(r) = (e^2 B^2/8m_r)r^2$, and (c) the potential due to the inhomogeneous in-plane electric field $V_E(x_r) = e \int F_x(x_r) dx_r$, where $x_r$ is the relative coordinate along $x$. Remarkably, the total potential exhibits two spatially separated potential wells and ensures that in the regime of intermediate field strengths, the wave function describing the relative electron-hole motion exhibits a large dipole moment well beyond what could be induced solely by the in-plane electric field \cite{Chestnov2021}.
%Importantly, the in-plane electric field $F_x$ shifts the potential minimum of the diamagnetic well away from $r = 0$, leading to a bi-modal form of the total potential. As noted in \cite{Chestnov2021}, in the regime of intermediate field strengths, this bimodal potential can lead to stretching of the relative wavefunction beyond what is expected from the dipole moment induced solely by the in-plane electric field. 
Our observations suggest that the role of an applied $B$-field is not limited to revealing the intrinsic properties of neutral excitons, but could instead be used as a tuning knob for dramatically modifying the excitonic wave function. Possible applications we foresee include enhancement of exciton-exciton interactions and the realization of effective gauge potentials for neutral excitons \cite{Chestnov2021, Togan2018}.
%The increased size of the magneto-electric exciton suggests that the role of an applied $B$-field is  may have important consequences for future work, especially for enhancing interactions and effecting gauge potentials.  

%To explain the origin of the larger diamagnetic shift of confined excitons, we note that in conventional systems such as semiconductor quantum wires and quantum dots, where exciton confinement \textcolor{black}{is directly linked to charge confinement}, the introduction of confinement always reduces the diamagnetic shift of excitons compared to the unconfined case. This can be simply understood by considering that confinement in these systems increases the electron-hole overlap and therefore reduces the rms size of excitons. In our case, however, the situation is the oppposite, i.e. the confinement applies only for the center-of-mass frame, whereas electrons and holes experience anti-confinement that pulls them apart from each other due to in-plane electric fields. The relative wavefunction is stretched along the confinement direction ($x$), leading to an elliptical shape. We can qualitatively describe this scenario by assigning unequal reduced masses $\mu_x$ and $\mu_y$ along the two directions, with $\mu_x < \mu_y$ due to the in-plane electric field. This leads to a reduction of the overall mass, which in turn leads to a larger diamagnetic shift compared to the 2D exciton. The observation of field-dependent diamagnetic shifts further highlights the unique nature of the confining potential in comparison to previously explored systems. 

In summary, by exploiting the ultra-strong exciton binding energy in monolayer MoSe$_2$ and repulsive exciton-charge interactions, we demonstrated quantum confinement of neutral excitons on nanoscopic length scales. The confinement takes place in a unique potential, which leads to dramatic modification of both center-of-mass and relative excitonic wave functions. Our method provides several crucial design advantages over material modulation approaches such as moir\'{e} potentials and strain engineering. These include: (i) deterministic positioning of tailor-made potentials by suitable design of electrodes; (ii) electrical tunability of exciton resonance energy which allows to overcome disorder and create multiple identical emitters; (iii) full quantum confinement without nanoscopic lithographic patterning, i.e. only by partial overlap of gates;  and (iv) quantum confinement of in-plane direct excitons, as opposed to layer-indirect excitons, which allows for strong coupling to light. 

We envision several exciting directions that are made possible by our work. First and foremost, strong confinement of excitons with a permanent dipole moment perpendicular to the wire axis is expected to strongly enhance exciton-exciton interactions \cite{Li2020,Kremser2020,Baek2020} while allowing for hybridization with a microcavity-mode \cite{Rosenberg2018,Togan2018}; consequently, we expect a 1D wire strongly coupled to a cavity mode to emerge as a building block of a strongly interacting photonic system \cite{Carusotto2013}. Even in the absence of cavity-coupling, strong interactions could enable the realization of an excitonic Tonks-Girardeau gas with photon correlations providing signatures of fermionization \cite{Rafal2021}. Last but not least, using proper design of electrodes, our method can be straightforwardly applied to achieve arbitrary confinement potential shapes, in particular lower dimensional quantum confined structures such as quantum dots or quantum rings and arrays thereof. \\

\noindent
\textbf{Acknowledgements} We thank A. Srivastava, I. Schwartz, R. Schmidt, A. Bergschneider, and N. Lassaline for insightful discussions. \textbf{Funding:} This work has been supported by Swiss National Science Foundation under grant 200021-178909/1. K.W. and T.T. acknowledge support from the Elemental Strategy Initiative conducted by MEXT, Japan, A3 Foresight by JSPS and CREST (grant number JPMJCR15F3) and JST. P.A.M. acknowledges funding from the European Union's Horizon 2020 program under Marie Sklodowska-Curie grant MSCA-IF-OptoTransport (843842). D.T. and D.J.N. acknowledge support by the Swiss National Science Foundation under grant 200021-165559. Synthesis of MoSe$_2$ crystals in Device 2 was supported by the United States National Science Foundation  Materials Research Science and Engineering Center DMR-2011738.
\textbf{Author contributions:} P.A.M, A.I. and D.T. conceptualized the project. D.T. and P.A.M. carried out the optics and transport experiments on Device 1. D.T. performed the electrostatic simulations with inputs from P.A.M. and A.P. D.T. fabricated Device 1 and X.L. fabricated Device 2. T.S. performed measurements on Device 2. A.P. assisted with measurements, device fabrication and simulations. K.W. and T.T. provided the h-BN crystals. S.L. and K.B. provided the MoSe$_2$ crystals for Device 2. M.K. and T.C. assisted P.A.M and D.T. with building the experimental setup. P.A.M, D.T. and A.I. wrote the manuscript. A.I., D.J.N., M.K., and P.A.M. supervised the project. \textbf{Competing interests:} The authors declare the following potential competing interests: D.T., P.A.M., M.K., A.P. and A.I. are seeking patent protection for ideas in this work. \textbf{Data availability:} The data that support the findings of this study will be made publicly available at the ETH Research Collection upon publication (\url{http://hdl.handle.net/20.500.11850/478320}).

%All authors participated in the discussion of results and preparation of the manuscript. \\ 

% \noindent
% \textbf{Supplementary Materials:}\\
% Supplementary information \\
% Table 1 \\
% Fig S1 - S9\\
% Movie S1 \\

\newpage
\bibliography{References/1Dexcitons_ref_v18.bib}

\begin{thebibliography}{40}%
\makeatletter
\providecommand \@ifxundefined [1]{%
 \@ifx{#1\undefined}
}%
\providecommand \@ifnum [1]{%
 \ifnum #1\expandafter \@firstoftwo
 \else \expandafter \@secondoftwo
 \fi
}%
\providecommand \@ifx [1]{%
 \ifx #1\expandafter \@firstoftwo
 \else \expandafter \@secondoftwo
 \fi
}%
\providecommand \natexlab [1]{#1}%
\providecommand \enquote  [1]{``#1''}%
\providecommand \bibnamefont  [1]{#1}%
\providecommand \bibfnamefont [1]{#1}%
\providecommand \citenamefont [1]{#1}%
\providecommand \href@noop [0]{\@secondoftwo}%
\providecommand \href [0]{\begingroup \@sanitize@url \@href}%
\providecommand \@href[1]{\@@startlink{#1}\@@href}%
\providecommand \@@href[1]{\endgroup#1\@@endlink}%
\providecommand \@sanitize@url [0]{\catcode `\\12\catcode `\$12\catcode
  `\&12\catcode `\#12\catcode `\^12\catcode `\_12\catcode `\%12\relax}%
\providecommand \@@startlink[1]{}%
\providecommand \@@endlink[0]{}%
\providecommand \url  [0]{\begingroup\@sanitize@url \@url }%
\providecommand \@url [1]{\endgroup\@href {#1}{\urlprefix }}%
\providecommand \urlprefix  [0]{URL }%
\providecommand \Eprint [0]{\href }%
\providecommand \doibase [0]{https://doi.org/}%
\providecommand \selectlanguage [0]{\@gobble}%
\providecommand \bibinfo  [0]{\@secondoftwo}%
\providecommand \bibfield  [0]{\@secondoftwo}%
\providecommand \translation [1]{[#1]}%
\providecommand \BibitemOpen [0]{}%
\providecommand \bibitemStop [0]{}%
\providecommand \bibitemNoStop [0]{.\EOS\space}%
\providecommand \EOS [0]{\spacefactor3000\relax}%
\providecommand \BibitemShut  [1]{\csname bibitem#1\endcsname}%
\let\auto@bib@innerbib\@empty
%</preamble>
\bibitem [{\citenamefont {Davies}(1997)}]{Davies1997}%
  \BibitemOpen
  \bibfield  {author} {\bibinfo {author} {\bibfnamefont {J.~H.}\ \bibnamefont
  {Davies}},\ }\href {https://doi.org/10.1017/CBO9780511819070} {\emph
  {\bibinfo {title} {{The Physics of Low-dimensional Semiconductors}}}}\
  (\bibinfo  {publisher} {Cambridge University Press},\ \bibinfo {year}
  {1997})\BibitemShut {NoStop}%
\bibitem [{\citenamefont {Hagn}\ \emph {et~al.}(1995)\citenamefont {Hagn},
  \citenamefont {Zrenner}, \citenamefont {B{\"{o}}hm},\ and\ \citenamefont
  {Weimann}}]{Hagn1995}%
  \BibitemOpen
  \bibfield  {author} {\bibinfo {author} {\bibfnamefont {M.}~\bibnamefont
  {Hagn}}, \bibinfo {author} {\bibfnamefont {A.}~\bibnamefont {Zrenner}},
  \bibinfo {author} {\bibfnamefont {G.}~\bibnamefont {B{\"{o}}hm}},\ and\
  \bibinfo {author} {\bibfnamefont {G.}~\bibnamefont {Weimann}},\ }\bibfield
  {title} {\bibinfo {title} {{Electric‐field‐induced exciton transport in
  coupled quantum well structures}},\ }\href {https://doi.org/10.1063/1.114677}
  {\bibfield  {journal} {\bibinfo  {journal} {Applied Physics Letters}\
  }\textbf {\bibinfo {volume} {67}},\ \bibinfo {pages} {232} (\bibinfo {year}
  {1995})}\BibitemShut {NoStop}%
\bibitem [{\citenamefont {Rapaport}\ \emph {et~al.}(2005)\citenamefont
  {Rapaport}, \citenamefont {Chen}, \citenamefont {Simon}, \citenamefont
  {Mitrofanov}, \citenamefont {Pfeiffer},\ and\ \citenamefont
  {Platzman}}]{Rapaport2005}%
  \BibitemOpen
  \bibfield  {author} {\bibinfo {author} {\bibfnamefont {R.}~\bibnamefont
  {Rapaport}}, \bibinfo {author} {\bibfnamefont {G.}~\bibnamefont {Chen}},
  \bibinfo {author} {\bibfnamefont {S.}~\bibnamefont {Simon}}, \bibinfo
  {author} {\bibfnamefont {O.}~\bibnamefont {Mitrofanov}}, \bibinfo {author}
  {\bibfnamefont {L.}~\bibnamefont {Pfeiffer}},\ and\ \bibinfo {author}
  {\bibfnamefont {P.~M.}\ \bibnamefont {Platzman}},\ }\bibfield  {title}
  {\bibinfo {title} {{Electrostatic traps for dipolar excitons}},\ }\href
  {https://doi.org/10.1103/PhysRevB.72.075428} {\bibfield  {journal} {\bibinfo
  {journal} {Physical Review B - Condensed Matter and Materials Physics}\
  }\textbf {\bibinfo {volume} {72}},\ \bibinfo {pages} {075428} (\bibinfo
  {year} {2005})}\BibitemShut {NoStop}%
\bibitem [{\citenamefont {G{\"{a}}rtner}\ \emph {et~al.}(2007)\citenamefont
  {G{\"{a}}rtner}, \citenamefont {Prechtel}, \citenamefont {Schuh},
  \citenamefont {Holleitner},\ and\ \citenamefont {Kotthaus}}]{Gartner2007}%
  \BibitemOpen
  \bibfield  {author} {\bibinfo {author} {\bibfnamefont {A.}~\bibnamefont
  {G{\"{a}}rtner}}, \bibinfo {author} {\bibfnamefont {L.}~\bibnamefont
  {Prechtel}}, \bibinfo {author} {\bibfnamefont {D.}~\bibnamefont {Schuh}},
  \bibinfo {author} {\bibfnamefont {A.~W.}\ \bibnamefont {Holleitner}},\ and\
  \bibinfo {author} {\bibfnamefont {J.~P.}\ \bibnamefont {Kotthaus}},\
  }\bibfield  {title} {\bibinfo {title} {{Micropatterned electrostatic traps
  for indirect excitons in coupled GaAs quantum wells}},\ }\href
  {https://doi.org/10.1103/PhysRevB.76.085304} {\bibfield  {journal} {\bibinfo
  {journal} {Physical Review B - Condensed Matter and Materials Physics}\
  }\textbf {\bibinfo {volume} {76}},\ \bibinfo {pages} {085304} (\bibinfo
  {year} {2007})}\BibitemShut {NoStop}%
\bibitem [{\citenamefont {Schinner}\ \emph {et~al.}(2013)\citenamefont
  {Schinner}, \citenamefont {Repp}, \citenamefont {Schubert}, \citenamefont
  {Rai}, \citenamefont {Reuter}, \citenamefont {Wieck}, \citenamefont
  {Govorov}, \citenamefont {Holleitner},\ and\ \citenamefont
  {Kotthaus}}]{Schinner2013}%
  \BibitemOpen
  \bibfield  {author} {\bibinfo {author} {\bibfnamefont {G.~J.}\ \bibnamefont
  {Schinner}}, \bibinfo {author} {\bibfnamefont {J.}~\bibnamefont {Repp}},
  \bibinfo {author} {\bibfnamefont {E.}~\bibnamefont {Schubert}}, \bibinfo
  {author} {\bibfnamefont {A.~K.}\ \bibnamefont {Rai}}, \bibinfo {author}
  {\bibfnamefont {D.}~\bibnamefont {Reuter}}, \bibinfo {author} {\bibfnamefont
  {A.~D.}\ \bibnamefont {Wieck}}, \bibinfo {author} {\bibfnamefont {A.~O.}\
  \bibnamefont {Govorov}}, \bibinfo {author} {\bibfnamefont {A.~W.}\
  \bibnamefont {Holleitner}},\ and\ \bibinfo {author} {\bibfnamefont {J.~P.}\
  \bibnamefont {Kotthaus}},\ }\bibfield  {title} {\bibinfo {title}
  {{Confinement and interaction of single indirect excitons in a
  voltage-controlled trap formed inside double InGaAs quantum wells}},\ }\href
  {https://doi.org/10.1103/PhysRevLett.110.127403} {\bibfield  {journal}
  {\bibinfo  {journal} {Physical Review Letters}\ }\textbf {\bibinfo {volume}
  {110}},\ \bibinfo {pages} {1} (\bibinfo {year} {2013})}\BibitemShut {NoStop}%
\bibitem [{\citenamefont {Butov}(2017)}]{Butov2017}%
  \BibitemOpen
  \bibfield  {author} {\bibinfo {author} {\bibfnamefont {L.~V.}\ \bibnamefont
  {Butov}},\ }\bibfield  {title} {\bibinfo {title} {{Excitonic devices}},\
  }\href {https://doi.org/10.1016/j.spmi.2016.12.035} {\bibfield  {journal}
  {\bibinfo  {journal} {Superlattices and Microstructures}\ }\textbf {\bibinfo
  {volume} {108}},\ \bibinfo {pages} {2} (\bibinfo {year} {2017})}\BibitemShut
  {NoStop}%
\bibitem [{\citenamefont {Hammack}\ \emph {et~al.}(2006)\citenamefont
  {Hammack}, \citenamefont {Gippius}, \citenamefont {Yang}, \citenamefont
  {Andreev}, \citenamefont {Butov}, \citenamefont {Hanson},\ and\ \citenamefont
  {Gossard}}]{Hammack2006}%
  \BibitemOpen
  \bibfield  {author} {\bibinfo {author} {\bibfnamefont {A.~T.}\ \bibnamefont
  {Hammack}}, \bibinfo {author} {\bibfnamefont {N.~A.}\ \bibnamefont
  {Gippius}}, \bibinfo {author} {\bibfnamefont {S.}~\bibnamefont {Yang}},
  \bibinfo {author} {\bibfnamefont {G.~O.}\ \bibnamefont {Andreev}}, \bibinfo
  {author} {\bibfnamefont {L.~V.}\ \bibnamefont {Butov}}, \bibinfo {author}
  {\bibfnamefont {M.}~\bibnamefont {Hanson}},\ and\ \bibinfo {author}
  {\bibfnamefont {A.~C.}\ \bibnamefont {Gossard}},\ }\bibfield  {title}
  {\bibinfo {title} {{Excitons in electrostatic traps}},\ }\href
  {https://doi.org/10.1063/1.2181276} {\bibfield  {journal} {\bibinfo
  {journal} {Journal of Applied Physics}\ }\textbf {\bibinfo {volume} {99}},\
  \bibinfo {pages} {066104} (\bibinfo {year} {2006})}\BibitemShut {NoStop}%
\bibitem [{\citenamefont {Unuchek}\ \emph {et~al.}(2018)\citenamefont
  {Unuchek}, \citenamefont {Ciarrocchi}, \citenamefont {Avsar}, \citenamefont
  {Watanabe}, \citenamefont {Taniguchi},\ and\ \citenamefont
  {Kis}}]{Unuchek2018}%
  \BibitemOpen
  \bibfield  {author} {\bibinfo {author} {\bibfnamefont {D.}~\bibnamefont
  {Unuchek}}, \bibinfo {author} {\bibfnamefont {A.}~\bibnamefont {Ciarrocchi}},
  \bibinfo {author} {\bibfnamefont {A.}~\bibnamefont {Avsar}}, \bibinfo
  {author} {\bibfnamefont {K.}~\bibnamefont {Watanabe}}, \bibinfo {author}
  {\bibfnamefont {T.}~\bibnamefont {Taniguchi}},\ and\ \bibinfo {author}
  {\bibfnamefont {A.}~\bibnamefont {Kis}},\ }\bibfield  {title} {\bibinfo
  {title} {{Room-temperature electrical control of exciton flux in a van der
  Waals heterostructure}},\ }\href {https://doi.org/10.1038/s41586-018-0357-y}
  {\bibfield  {journal} {\bibinfo  {journal} {Nature}\ }\textbf {\bibinfo
  {volume} {560}},\ \bibinfo {pages} {340} (\bibinfo {year}
  {2018})}\BibitemShut {NoStop}%
\bibitem [{\citenamefont {Wang}\ \emph {et~al.}(2018)\citenamefont {Wang},
  \citenamefont {Chernikov}, \citenamefont {Glazov}, \citenamefont {Heinz},
  \citenamefont {Marie}, \citenamefont {Amand},\ and\ \citenamefont
  {Urbaszek}}]{Wang2018a}%
  \BibitemOpen
  \bibfield  {author} {\bibinfo {author} {\bibfnamefont {G.}~\bibnamefont
  {Wang}}, \bibinfo {author} {\bibfnamefont {A.}~\bibnamefont {Chernikov}},
  \bibinfo {author} {\bibfnamefont {M.~M.}\ \bibnamefont {Glazov}}, \bibinfo
  {author} {\bibfnamefont {T.~F.}\ \bibnamefont {Heinz}}, \bibinfo {author}
  {\bibfnamefont {X.}~\bibnamefont {Marie}}, \bibinfo {author} {\bibfnamefont
  {T.}~\bibnamefont {Amand}},\ and\ \bibinfo {author} {\bibfnamefont
  {B.}~\bibnamefont {Urbaszek}},\ }\bibfield  {title} {\bibinfo {title}
  {{Colloquium: Excitons in atomically thin transition metal
  dichalcogenides}},\ }\href {https://doi.org/10.1103/RevModPhys.90.021001}
  {\bibfield  {journal} {\bibinfo  {journal} {Reviews of Modern Physics}\
  }\textbf {\bibinfo {volume} {90}},\ \bibinfo {pages} {021001} (\bibinfo
  {year} {2018})}\BibitemShut {NoStop}%
\bibitem [{\citenamefont {Liu}\ \emph {et~al.}(2020)\citenamefont {Liu},
  \citenamefont {Dini}, \citenamefont {Tan}, \citenamefont {Liew},
  \citenamefont {Novoselov},\ and\ \citenamefont {Gao}}]{Liu2020}%
  \BibitemOpen
  \bibfield  {author} {\bibinfo {author} {\bibfnamefont {Y.}~\bibnamefont
  {Liu}}, \bibinfo {author} {\bibfnamefont {K.}~\bibnamefont {Dini}}, \bibinfo
  {author} {\bibfnamefont {Q.}~\bibnamefont {Tan}}, \bibinfo {author}
  {\bibfnamefont {T.}~\bibnamefont {Liew}}, \bibinfo {author} {\bibfnamefont
  {K.~S.}\ \bibnamefont {Novoselov}},\ and\ \bibinfo {author} {\bibfnamefont
  {W.}~\bibnamefont {Gao}},\ }\bibfield  {title} {\bibinfo {title}
  {{Electrically controllable router of interlayer excitons}},\ }\bibfield
  {journal} {\bibinfo  {journal} {Science Advances}\ }\textbf {\bibinfo
  {volume} {6}},\ \href {https://doi.org/10.1126/sciadv.aba1830}
  {10.1126/sciadv.aba1830} (\bibinfo {year} {2020})\BibitemShut {NoStop}%
\bibitem [{\citenamefont {Jauregui}\ \emph {et~al.}(2019)\citenamefont
  {Jauregui}, \citenamefont {Joe}, \citenamefont {Pistunova}, \citenamefont
  {Wild}, \citenamefont {High}, \citenamefont {Zhou}, \citenamefont {Scuri},
  \citenamefont {de~Greve}, \citenamefont {Sushko}, \citenamefont {Yu},
  \citenamefont {Taniguchi}, \citenamefont {Watanabe}, \citenamefont
  {Needleman}, \citenamefont {Lukin}, \citenamefont {Park},\ and\ \citenamefont
  {Kim}}]{Jauregui2019}%
  \BibitemOpen
  \bibfield  {author} {\bibinfo {author} {\bibfnamefont {L.~A.}\ \bibnamefont
  {Jauregui}}, \bibinfo {author} {\bibfnamefont {A.~Y.}\ \bibnamefont {Joe}},
  \bibinfo {author} {\bibfnamefont {K.}~\bibnamefont {Pistunova}}, \bibinfo
  {author} {\bibfnamefont {D.~S.}\ \bibnamefont {Wild}}, \bibinfo {author}
  {\bibfnamefont {A.~A.}\ \bibnamefont {High}}, \bibinfo {author}
  {\bibfnamefont {Y.}~\bibnamefont {Zhou}}, \bibinfo {author} {\bibfnamefont
  {G.}~\bibnamefont {Scuri}}, \bibinfo {author} {\bibfnamefont
  {K.}~\bibnamefont {de~Greve}}, \bibinfo {author} {\bibfnamefont
  {A.}~\bibnamefont {Sushko}}, \bibinfo {author} {\bibfnamefont {C.~H.}\
  \bibnamefont {Yu}}, \bibinfo {author} {\bibfnamefont {T.}~\bibnamefont
  {Taniguchi}}, \bibinfo {author} {\bibfnamefont {K.}~\bibnamefont {Watanabe}},
  \bibinfo {author} {\bibfnamefont {D.~J.}\ \bibnamefont {Needleman}}, \bibinfo
  {author} {\bibfnamefont {M.~D.}\ \bibnamefont {Lukin}}, \bibinfo {author}
  {\bibfnamefont {H.}~\bibnamefont {Park}},\ and\ \bibinfo {author}
  {\bibfnamefont {P.}~\bibnamefont {Kim}},\ }\bibfield  {title} {\bibinfo
  {title} {{Electrical control of interlayer exciton dynamics in atomically
  thin heterostructures}},\ }\href {https://doi.org/10.1126/science.aaw4194}
  {\bibfield  {journal} {\bibinfo  {journal} {Science}\ }\textbf {\bibinfo
  {volume} {366}},\ \bibinfo {pages} {870} (\bibinfo {year} {2019})},\ \Eprint
  {https://arxiv.org/abs/1812.08691} {arXiv:1812.08691} \BibitemShut {NoStop}%
\bibitem [{\citenamefont {Cavalcante}\ \emph {et~al.}(2018)\citenamefont
  {Cavalcante}, \citenamefont {{Da Costa}}, \citenamefont {Farias},
  \citenamefont {Reichman},\ and\ \citenamefont {Chaves}}]{Cavalcante2018}%
  \BibitemOpen
  \bibfield  {author} {\bibinfo {author} {\bibfnamefont {L.~S.}\ \bibnamefont
  {Cavalcante}}, \bibinfo {author} {\bibfnamefont {D.~R.}\ \bibnamefont {{Da
  Costa}}}, \bibinfo {author} {\bibfnamefont {G.~A.}\ \bibnamefont {Farias}},
  \bibinfo {author} {\bibfnamefont {D.~R.}\ \bibnamefont {Reichman}},\ and\
  \bibinfo {author} {\bibfnamefont {A.}~\bibnamefont {Chaves}},\ }\bibfield
  {title} {\bibinfo {title} {{Stark shift of excitons and trions in
  two-dimensional materials}},\ }\href
  {https://doi.org/10.1103/PhysRevB.98.245309} {\bibfield  {journal} {\bibinfo
  {journal} {Physical Review B}\ }\textbf {\bibinfo {volume} {98}},\ \bibinfo
  {pages} {245309} (\bibinfo {year} {2018})}\BibitemShut {NoStop}%
\bibitem [{\citenamefont {Efimkin}\ and\ \citenamefont
  {MacDonald}(2017)}]{Efimkin2017}%
  \BibitemOpen
  \bibfield  {author} {\bibinfo {author} {\bibfnamefont {D.~K.}\ \bibnamefont
  {Efimkin}}\ and\ \bibinfo {author} {\bibfnamefont {A.~H.}\ \bibnamefont
  {MacDonald}},\ }\bibfield  {title} {\bibinfo {title} {{Many-body theory of
  trion absorption features in two-dimensional semiconductors}},\ }\href
  {https://doi.org/10.1103/PhysRevB.95.035417} {\bibfield  {journal} {\bibinfo
  {journal} {Physical Review B}\ }\textbf {\bibinfo {volume} {95}},\ \bibinfo
  {pages} {035417} (\bibinfo {year} {2017})}\BibitemShut {NoStop}%
\bibitem [{\citenamefont {Sidler}\ \emph {et~al.}(2017)\citenamefont {Sidler},
  \citenamefont {Back}, \citenamefont {Cotlet}, \citenamefont {Srivastava},
  \citenamefont {Fink}, \citenamefont {Kroner}, \citenamefont {Demler},\ and\
  \citenamefont {Imamoglu}}]{Sidler2017}%
  \BibitemOpen
  \bibfield  {author} {\bibinfo {author} {\bibfnamefont {M.}~\bibnamefont
  {Sidler}}, \bibinfo {author} {\bibfnamefont {P.}~\bibnamefont {Back}},
  \bibinfo {author} {\bibfnamefont {O.}~\bibnamefont {Cotlet}}, \bibinfo
  {author} {\bibfnamefont {A.}~\bibnamefont {Srivastava}}, \bibinfo {author}
  {\bibfnamefont {T.}~\bibnamefont {Fink}}, \bibinfo {author} {\bibfnamefont
  {M.}~\bibnamefont {Kroner}}, \bibinfo {author} {\bibfnamefont
  {E.}~\bibnamefont {Demler}},\ and\ \bibinfo {author} {\bibfnamefont
  {A.}~\bibnamefont {Imamoglu}},\ }\bibfield  {title} {\bibinfo {title} {{Fermi
  polaron-polaritons in charge-tunable atomically thin semiconductors}},\
  }\href {https://doi.org/10.1038/nphys3949} {\bibfield  {journal} {\bibinfo
  {journal} {Nature Physics}\ }\textbf {\bibinfo {volume} {13}},\ \bibinfo
  {pages} {255} (\bibinfo {year} {2017})},\ \Eprint
  {https://arxiv.org/abs/1603.09215} {arXiv:1603.09215} \BibitemShut {NoStop}%
\bibitem [{\citenamefont {Chervy}\ \emph {et~al.}(2020)\citenamefont {Chervy},
  \citenamefont {Kn{\"{u}}ppel}, \citenamefont {Abbaspour}, \citenamefont
  {Lupatini}, \citenamefont {F{\"{a}}lt}, \citenamefont {Wegscheider},
  \citenamefont {Kroner},\ and\ \citenamefont {Imamoǧlu}}]{Chervy2020}%
  \BibitemOpen
  \bibfield  {author} {\bibinfo {author} {\bibfnamefont {T.}~\bibnamefont
  {Chervy}}, \bibinfo {author} {\bibfnamefont {P.}~\bibnamefont
  {Kn{\"{u}}ppel}}, \bibinfo {author} {\bibfnamefont {H.}~\bibnamefont
  {Abbaspour}}, \bibinfo {author} {\bibfnamefont {M.}~\bibnamefont {Lupatini}},
  \bibinfo {author} {\bibfnamefont {S.}~\bibnamefont {F{\"{a}}lt}}, \bibinfo
  {author} {\bibfnamefont {W.}~\bibnamefont {Wegscheider}}, \bibinfo {author}
  {\bibfnamefont {M.}~\bibnamefont {Kroner}},\ and\ \bibinfo {author}
  {\bibfnamefont {A.}~\bibnamefont {Imamoǧlu}},\ }\bibfield  {title} {\bibinfo
  {title} {{Accelerating Polaritons with External Electric and Magnetic
  Fields}},\ }\href {https://doi.org/10.1103/PhysRevX.10.011040} {\bibfield
  {journal} {\bibinfo  {journal} {Physical Review X}\ }\textbf {\bibinfo
  {volume} {10}},\ \bibinfo {pages} {011040} (\bibinfo {year}
  {2020})}\BibitemShut {NoStop}%
\bibitem [{\citenamefont {Larentis}\ \emph {et~al.}(2018)\citenamefont
  {Larentis}, \citenamefont {Movva}, \citenamefont {Fallahazad}, \citenamefont
  {Kim}, \citenamefont {Behroozi}, \citenamefont {Taniguchi}, \citenamefont
  {Watanabe}, \citenamefont {Banerjee},\ and\ \citenamefont
  {Tutuc}}]{Larentis2018}%
  \BibitemOpen
  \bibfield  {author} {\bibinfo {author} {\bibfnamefont {S.}~\bibnamefont
  {Larentis}}, \bibinfo {author} {\bibfnamefont {H.~C.~P.}\ \bibnamefont
  {Movva}}, \bibinfo {author} {\bibfnamefont {B.}~\bibnamefont {Fallahazad}},
  \bibinfo {author} {\bibfnamefont {K.}~\bibnamefont {Kim}}, \bibinfo {author}
  {\bibfnamefont {A.}~\bibnamefont {Behroozi}}, \bibinfo {author}
  {\bibfnamefont {T.}~\bibnamefont {Taniguchi}}, \bibinfo {author}
  {\bibfnamefont {K.}~\bibnamefont {Watanabe}}, \bibinfo {author}
  {\bibfnamefont {S.~K.}\ \bibnamefont {Banerjee}},\ and\ \bibinfo {author}
  {\bibfnamefont {E.}~\bibnamefont {Tutuc}},\ }\bibfield  {title} {\bibinfo
  {title} {{Large effective mass and interaction-enhanced Zeeman splitting of
  K-valley electrons in {MoSe$_2$}}},\ }\href
  {https://doi.org/10.1103/PhysRevB.97.201407} {\bibfield  {journal} {\bibinfo
  {journal} {Physical Review B}\ }\textbf {\bibinfo {volume} {97}},\ \bibinfo
  {pages} {201407} (\bibinfo {year} {2018})}\BibitemShut {NoStop}%
\bibitem [{\citenamefont {Zhang}\ \emph {et~al.}(2014)\citenamefont {Zhang},
  \citenamefont {Chang}, \citenamefont {Zhou}, \citenamefont {Cui},
  \citenamefont {Yan}, \citenamefont {Liu}, \citenamefont {Schmitt},
  \citenamefont {Lee}, \citenamefont {Moore}, \citenamefont {Chen},
  \citenamefont {Lin}, \citenamefont {Jeng}, \citenamefont {Mo}, \citenamefont
  {Hussain}, \citenamefont {Bansil},\ and\ \citenamefont {Shen}}]{Zhang2014}%
  \BibitemOpen
  \bibfield  {author} {\bibinfo {author} {\bibfnamefont {Y.}~\bibnamefont
  {Zhang}}, \bibinfo {author} {\bibfnamefont {T.-R.}\ \bibnamefont {Chang}},
  \bibinfo {author} {\bibfnamefont {B.}~\bibnamefont {Zhou}}, \bibinfo {author}
  {\bibfnamefont {Y.-T.}\ \bibnamefont {Cui}}, \bibinfo {author} {\bibfnamefont
  {H.}~\bibnamefont {Yan}}, \bibinfo {author} {\bibfnamefont {Z.}~\bibnamefont
  {Liu}}, \bibinfo {author} {\bibfnamefont {F.}~\bibnamefont {Schmitt}},
  \bibinfo {author} {\bibfnamefont {J.}~\bibnamefont {Lee}}, \bibinfo {author}
  {\bibfnamefont {R.}~\bibnamefont {Moore}}, \bibinfo {author} {\bibfnamefont
  {Y.}~\bibnamefont {Chen}}, \bibinfo {author} {\bibfnamefont {H.}~\bibnamefont
  {Lin}}, \bibinfo {author} {\bibfnamefont {H.-T.}\ \bibnamefont {Jeng}},
  \bibinfo {author} {\bibfnamefont {S.-K.}\ \bibnamefont {Mo}}, \bibinfo
  {author} {\bibfnamefont {Z.}~\bibnamefont {Hussain}}, \bibinfo {author}
  {\bibfnamefont {A.}~\bibnamefont {Bansil}},\ and\ \bibinfo {author}
  {\bibfnamefont {Z.-X.}\ \bibnamefont {Shen}},\ }\bibfield  {title} {\bibinfo
  {title} {{Direct observation of the transition from indirect to direct
  bandgap in atomically thin epitaxial {MoSe$_2$}}},\ }\href
  {https://doi.org/10.1038/nnano.2013.277} {\bibfield  {journal} {\bibinfo
  {journal} {Nature Nanotechnology}\ }\textbf {\bibinfo {volume} {9}},\
  \bibinfo {pages} {111} (\bibinfo {year} {2014})}\BibitemShut {NoStop}%
\bibitem [{\citenamefont {Goryca}\ \emph {et~al.}(2019)\citenamefont {Goryca},
  \citenamefont {Li}, \citenamefont {Stier}, \citenamefont {Taniguchi},
  \citenamefont {Watanabe}, \citenamefont {Courtade}, \citenamefont {Shree},
  \citenamefont {Robert}, \citenamefont {Urbaszek}, \citenamefont {Marie},\
  and\ \citenamefont {Crooker}}]{Goryca2019}%
  \BibitemOpen
  \bibfield  {author} {\bibinfo {author} {\bibfnamefont {M.}~\bibnamefont
  {Goryca}}, \bibinfo {author} {\bibfnamefont {J.}~\bibnamefont {Li}}, \bibinfo
  {author} {\bibfnamefont {A.~V.}\ \bibnamefont {Stier}}, \bibinfo {author}
  {\bibfnamefont {T.}~\bibnamefont {Taniguchi}}, \bibinfo {author}
  {\bibfnamefont {K.}~\bibnamefont {Watanabe}}, \bibinfo {author}
  {\bibfnamefont {E.}~\bibnamefont {Courtade}}, \bibinfo {author}
  {\bibfnamefont {S.}~\bibnamefont {Shree}}, \bibinfo {author} {\bibfnamefont
  {C.}~\bibnamefont {Robert}}, \bibinfo {author} {\bibfnamefont
  {B.}~\bibnamefont {Urbaszek}}, \bibinfo {author} {\bibfnamefont
  {X.}~\bibnamefont {Marie}},\ and\ \bibinfo {author} {\bibfnamefont {S.~A.}\
  \bibnamefont {Crooker}},\ }\bibfield  {title} {\bibinfo {title} {{Revealing
  exciton masses and dielectric properties of monolayer semiconductors with
  high magnetic fields}},\ }\href {https://doi.org/10.1038/s41467-019-12180-y}
  {\bibfield  {journal} {\bibinfo  {journal} {Nature Communications}\ }\textbf
  {\bibinfo {volume} {10}},\ \bibinfo {pages} {4172} (\bibinfo {year}
  {2019})}\BibitemShut {NoStop}%
\bibitem [{som()}]{som}%
  \BibitemOpen
  \href@noop {} {\bibinfo {title} {{See supplementary materials}}}\BibitemShut
  {NoStop}%
\bibitem [{\citenamefont {Wang}(2001)}]{Wang2001}%
  \BibitemOpen
  \bibfield  {author} {\bibinfo {author} {\bibfnamefont {J.}~\bibnamefont
  {Wang}},\ }\bibfield  {title} {\bibinfo {title} {{Highly Polarized
  Photoluminescence and Photodetection from Single Indium Phosphide
  Nanowires}},\ }\href {https://doi.org/10.1126/science.1062340} {\bibfield
  {journal} {\bibinfo  {journal} {Science}\ }\textbf {\bibinfo {volume}
  {293}},\ \bibinfo {pages} {1455} (\bibinfo {year} {2001})}\BibitemShut
  {NoStop}%
\bibitem [{\citenamefont {Akiyama}\ \emph {et~al.}(1996)\citenamefont
  {Akiyama}, \citenamefont {Someya},\ and\ \citenamefont
  {Sakaki}}]{Akiyama1996}%
  \BibitemOpen
  \bibfield  {author} {\bibinfo {author} {\bibfnamefont {H.}~\bibnamefont
  {Akiyama}}, \bibinfo {author} {\bibfnamefont {T.}~\bibnamefont {Someya}},\
  and\ \bibinfo {author} {\bibfnamefont {H.}~\bibnamefont {Sakaki}},\
  }\bibfield  {title} {\bibinfo {title} {{Optical anisotropy in 5-nm-scale
  T-shaped quantum wires fabricated by the cleaved-edge overgrowth method}},\
  }\href {https://doi.org/10.1103/PhysRevB.53.R4229} {\bibfield  {journal}
  {\bibinfo  {journal} {Physical Review B}\ }\textbf {\bibinfo {volume} {53}},\
  \bibinfo {pages} {R4229} (\bibinfo {year} {1996})}\BibitemShut {NoStop}%
\bibitem [{\citenamefont {Lefebvre}\ \emph {et~al.}(2004)\citenamefont
  {Lefebvre}, \citenamefont {Fraser}, \citenamefont {Finnie},\ and\
  \citenamefont {Homma}}]{Lefebvre2004}%
  \BibitemOpen
  \bibfield  {author} {\bibinfo {author} {\bibfnamefont {J.}~\bibnamefont
  {Lefebvre}}, \bibinfo {author} {\bibfnamefont {J.~M.}\ \bibnamefont
  {Fraser}}, \bibinfo {author} {\bibfnamefont {P.}~\bibnamefont {Finnie}},\
  and\ \bibinfo {author} {\bibfnamefont {Y.}~\bibnamefont {Homma}},\ }\bibfield
   {title} {\bibinfo {title} {{Photoluminescence from an individual
  single-walled carbon nanotube}},\ }\href
  {https://doi.org/10.1103/PhysRevB.69.075403} {\bibfield  {journal} {\bibinfo
  {journal} {Physical Review B}\ }\textbf {\bibinfo {volume} {69}},\ \bibinfo
  {pages} {075403} (\bibinfo {year} {2004})}\BibitemShut {NoStop}%
\bibitem [{\citenamefont {Bai}\ \emph {et~al.}(2020)\citenamefont {Bai},
  \citenamefont {Zhou}, \citenamefont {Wang}, \citenamefont {Wu}, \citenamefont
  {McGilly}, \citenamefont {Halbertal}, \citenamefont {Lo}, \citenamefont
  {Liu}, \citenamefont {Ardelean}, \citenamefont {Rivera}, \citenamefont
  {Finney}, \citenamefont {Yang}, \citenamefont {Basov}, \citenamefont {Yao},
  \citenamefont {Xu}, \citenamefont {Hone}, \citenamefont {Pasupathy},\ and\
  \citenamefont {Zhu}}]{Bai2020}%
  \BibitemOpen
  \bibfield  {author} {\bibinfo {author} {\bibfnamefont {Y.}~\bibnamefont
  {Bai}}, \bibinfo {author} {\bibfnamefont {L.}~\bibnamefont {Zhou}}, \bibinfo
  {author} {\bibfnamefont {J.}~\bibnamefont {Wang}}, \bibinfo {author}
  {\bibfnamefont {W.}~\bibnamefont {Wu}}, \bibinfo {author} {\bibfnamefont
  {L.~J.}\ \bibnamefont {McGilly}}, \bibinfo {author} {\bibfnamefont
  {D.}~\bibnamefont {Halbertal}}, \bibinfo {author} {\bibfnamefont {C.~F.~B.}\
  \bibnamefont {Lo}}, \bibinfo {author} {\bibfnamefont {F.}~\bibnamefont
  {Liu}}, \bibinfo {author} {\bibfnamefont {J.}~\bibnamefont {Ardelean}},
  \bibinfo {author} {\bibfnamefont {P.}~\bibnamefont {Rivera}}, \bibinfo
  {author} {\bibfnamefont {N.~R.}\ \bibnamefont {Finney}}, \bibinfo {author}
  {\bibfnamefont {X.~C.}\ \bibnamefont {Yang}}, \bibinfo {author}
  {\bibfnamefont {D.~N.}\ \bibnamefont {Basov}}, \bibinfo {author}
  {\bibfnamefont {W.}~\bibnamefont {Yao}}, \bibinfo {author} {\bibfnamefont
  {X.}~\bibnamefont {Xu}}, \bibinfo {author} {\bibfnamefont {J.}~\bibnamefont
  {Hone}}, \bibinfo {author} {\bibfnamefont {A.~N.}\ \bibnamefont
  {Pasupathy}},\ and\ \bibinfo {author} {\bibfnamefont {X.~Y.}\ \bibnamefont
  {Zhu}},\ }\bibfield  {title} {\bibinfo {title} {{Excitons in strain-induced
  one-dimensional moir{\'{e}} potentials at transition metal dichalcogenide
  heterojunctions}},\ }\href {https://doi.org/10.1038/s41563-020-0730-8}
  {\bibfield  {journal} {\bibinfo  {journal} {Nature Materials}\ }\textbf
  {\bibinfo {volume} {19}},\ \bibinfo {pages} {1068} (\bibinfo {year}
  {2020})}\BibitemShut {NoStop}%
\bibitem [{\citenamefont {Glazov}\ \emph {et~al.}(2015)\citenamefont {Glazov},
  \citenamefont {Ivchenko}, \citenamefont {Wang}, \citenamefont {Amand},
  \citenamefont {Marie}, \citenamefont {Urbaszek},\ and\ \citenamefont
  {Liu}}]{Glazov2015}%
  \BibitemOpen
  \bibfield  {author} {\bibinfo {author} {\bibfnamefont {M.~M.}\ \bibnamefont
  {Glazov}}, \bibinfo {author} {\bibfnamefont {E.~L.}\ \bibnamefont
  {Ivchenko}}, \bibinfo {author} {\bibfnamefont {G.}~\bibnamefont {Wang}},
  \bibinfo {author} {\bibfnamefont {T.}~\bibnamefont {Amand}}, \bibinfo
  {author} {\bibfnamefont {X.}~\bibnamefont {Marie}}, \bibinfo {author}
  {\bibfnamefont {B.}~\bibnamefont {Urbaszek}},\ and\ \bibinfo {author}
  {\bibfnamefont {B.~L.}\ \bibnamefont {Liu}},\ }\bibfield  {title} {\bibinfo
  {title} {{Spin and valley dynamics of excitons in transition metal
  dichalcogenide monolayers}},\ }\href {https://doi.org/10.1002/pssb.201552211}
  {\bibfield  {journal} {\bibinfo  {journal} {physica status solidi (b)}\
  }\textbf {\bibinfo {volume} {252}},\ \bibinfo {pages} {2349} (\bibinfo {year}
  {2015})}\BibitemShut {NoStop}%
\bibitem [{\citenamefont {Yu}\ \emph {et~al.}(2015)\citenamefont {Yu},
  \citenamefont {Cui}, \citenamefont {Xu},\ and\ \citenamefont {Yao}}]{Yu2015}%
  \BibitemOpen
  \bibfield  {author} {\bibinfo {author} {\bibfnamefont {H.}~\bibnamefont
  {Yu}}, \bibinfo {author} {\bibfnamefont {X.}~\bibnamefont {Cui}}, \bibinfo
  {author} {\bibfnamefont {X.}~\bibnamefont {Xu}},\ and\ \bibinfo {author}
  {\bibfnamefont {W.}~\bibnamefont {Yao}},\ }\bibfield  {title} {\bibinfo
  {title} {{Valley excitons in two-dimensional semiconductors}},\ }\href
  {https://doi.org/10.1093/nsr/nwu078} {\bibfield  {journal} {\bibinfo
  {journal} {National Science Review}\ }\textbf {\bibinfo {volume} {2}},\
  \bibinfo {pages} {57} (\bibinfo {year} {2015})}\BibitemShut {NoStop}%
\bibitem [{\citenamefont {Stier}\ \emph {et~al.}(2016)\citenamefont {Stier},
  \citenamefont {McCreary}, \citenamefont {Jonker},\ and\ \citenamefont
  {Crooker}}]{Stier2016}%
  \BibitemOpen
  \bibfield  {author} {\bibinfo {author} {\bibfnamefont {A.~V.}\ \bibnamefont
  {Stier}}, \bibinfo {author} {\bibfnamefont {K.~M.}\ \bibnamefont {McCreary}},
  \bibinfo {author} {\bibfnamefont {B.~T.}\ \bibnamefont {Jonker}},\ and\
  \bibinfo {author} {\bibfnamefont {S.}~\bibnamefont {Crooker}},\ }\bibfield
  {title} {\bibinfo {title} {{Exciton diamagnetic shifts and valley Zeeman
  effects in monolayer {WS$_2$} and {MoS$_2$} to 65 Tesla}},\ }\bibfield
  {journal} {\bibinfo  {journal} {Nature Communications}\ }\textbf {\bibinfo
  {volume} {7}},\ \href {https://doi.org/10.1038/ncomms10643}
  {10.1038/ncomms10643} (\bibinfo {year} {2016})\BibitemShut {NoStop}%
\bibitem [{\citenamefont {Chestnov}\ \emph {et~al.}(2021)\citenamefont
  {Chestnov}, \citenamefont {Arakelian},\ and\ \citenamefont
  {Kavokin}}]{Chestnov2021}%
  \BibitemOpen
  \bibfield  {author} {\bibinfo {author} {\bibfnamefont {I.~Y.}\ \bibnamefont
  {Chestnov}}, \bibinfo {author} {\bibfnamefont {S.~M.}\ \bibnamefont
  {Arakelian}},\ and\ \bibinfo {author} {\bibfnamefont {A.~V.}\ \bibnamefont
  {Kavokin}},\ }\bibfield  {title} {\bibinfo {title} {Giant synthetic gauge
  field for spinless microcavity polaritons in crossed electric and magnetic
  fields},\ }\href {https://doi.org/10.1088/1367-2630/abe2bf} {\bibfield
  {journal} {\bibinfo  {journal} {New Journal of Physics}\ }\textbf {\bibinfo
  {volume} {23}},\ \bibinfo {pages} {023024} (\bibinfo {year}
  {2021})}\BibitemShut {NoStop}%
\bibitem [{\citenamefont {Togan}\ \emph {et~al.}(2018)\citenamefont {Togan},
  \citenamefont {Lim}, \citenamefont {Faelt}, \citenamefont {Wegscheider},\
  and\ \citenamefont {Imamoglu}}]{Togan2018}%
  \BibitemOpen
  \bibfield  {author} {\bibinfo {author} {\bibfnamefont {E.}~\bibnamefont
  {Togan}}, \bibinfo {author} {\bibfnamefont {H.-T.}\ \bibnamefont {Lim}},
  \bibinfo {author} {\bibfnamefont {S.}~\bibnamefont {Faelt}}, \bibinfo
  {author} {\bibfnamefont {W.}~\bibnamefont {Wegscheider}},\ and\ \bibinfo
  {author} {\bibfnamefont {A.}~\bibnamefont {Imamoglu}},\ }\bibfield  {title}
  {\bibinfo {title} {{Enhanced Interactions between Dipolar Polaritons}},\
  }\href {https://doi.org/10.1103/PhysRevLett.121.227402} {\bibfield  {journal}
  {\bibinfo  {journal} {Physical Review Letters}\ }\textbf {\bibinfo {volume}
  {121}},\ \bibinfo {pages} {227402} (\bibinfo {year} {2018})}\BibitemShut
  {NoStop}%
\bibitem [{\citenamefont {Li}\ \emph {et~al.}(2020)\citenamefont {Li},
  \citenamefont {Lu}, \citenamefont {Dubey}, \citenamefont {Devenica},\ and\
  \citenamefont {Srivastava}}]{Li2020}%
  \BibitemOpen
  \bibfield  {author} {\bibinfo {author} {\bibfnamefont {W.}~\bibnamefont
  {Li}}, \bibinfo {author} {\bibfnamefont {X.}~\bibnamefont {Lu}}, \bibinfo
  {author} {\bibfnamefont {S.}~\bibnamefont {Dubey}}, \bibinfo {author}
  {\bibfnamefont {L.}~\bibnamefont {Devenica}},\ and\ \bibinfo {author}
  {\bibfnamefont {A.}~\bibnamefont {Srivastava}},\ }\bibfield  {title}
  {\bibinfo {title} {{Dipolar interactions between localized interlayer
  excitons in van der Waals heterostructures}},\ }\href
  {https://doi.org/10.1038/s41563-020-0661-4} {\bibfield  {journal} {\bibinfo
  {journal} {Nature Materials}\ }\textbf {\bibinfo {volume} {19}},\ \bibinfo
  {pages} {624} (\bibinfo {year} {2020})}\BibitemShut {NoStop}%
\bibitem [{\citenamefont {Kremser}\ \emph {et~al.}(2020)\citenamefont
  {Kremser}, \citenamefont {Brotons-Gisbert}, \citenamefont {Kn{\"{o}}rzer},
  \citenamefont {G{\"{u}}ckelhorn}, \citenamefont {Meyer}, \citenamefont
  {Barbone}, \citenamefont {Stier}, \citenamefont {Gerardot}, \citenamefont
  {M{\"{u}}ller},\ and\ \citenamefont {Finley}}]{Kremser2020}%
  \BibitemOpen
  \bibfield  {author} {\bibinfo {author} {\bibfnamefont {M.}~\bibnamefont
  {Kremser}}, \bibinfo {author} {\bibfnamefont {M.}~\bibnamefont
  {Brotons-Gisbert}}, \bibinfo {author} {\bibfnamefont {J.}~\bibnamefont
  {Kn{\"{o}}rzer}}, \bibinfo {author} {\bibfnamefont {J.}~\bibnamefont
  {G{\"{u}}ckelhorn}}, \bibinfo {author} {\bibfnamefont {M.}~\bibnamefont
  {Meyer}}, \bibinfo {author} {\bibfnamefont {M.}~\bibnamefont {Barbone}},
  \bibinfo {author} {\bibfnamefont {A.~V.}\ \bibnamefont {Stier}}, \bibinfo
  {author} {\bibfnamefont {B.~D.}\ \bibnamefont {Gerardot}}, \bibinfo {author}
  {\bibfnamefont {K.}~\bibnamefont {M{\"{u}}ller}},\ and\ \bibinfo {author}
  {\bibfnamefont {J.~J.}\ \bibnamefont {Finley}},\ }\bibfield  {title}
  {\bibinfo {title} {{Discrete interactions between a few interlayer excitons
  trapped at a {MoSe$_2$}–{WSe$_2$} heterointerface}},\ }\href
  {https://doi.org/10.1038/s41699-020-0141-3} {\bibfield  {journal} {\bibinfo
  {journal} {npj 2D Materials and Applications}\ }\textbf {\bibinfo {volume}
  {4}},\ \bibinfo {pages} {8} (\bibinfo {year} {2020})}\BibitemShut {NoStop}%
\bibitem [{\citenamefont {Baek}\ \emph {et~al.}(2020)\citenamefont {Baek},
  \citenamefont {Brotons-Gisbert}, \citenamefont {Koong}, \citenamefont
  {Campbell}, \citenamefont {Rambach}, \citenamefont {Watanabe}, \citenamefont
  {Taniguchi},\ and\ \citenamefont {Gerardot}}]{Baek2020}%
  \BibitemOpen
  \bibfield  {author} {\bibinfo {author} {\bibfnamefont {H.}~\bibnamefont
  {Baek}}, \bibinfo {author} {\bibfnamefont {M.}~\bibnamefont
  {Brotons-Gisbert}}, \bibinfo {author} {\bibfnamefont {Z.~X.}\ \bibnamefont
  {Koong}}, \bibinfo {author} {\bibfnamefont {A.}~\bibnamefont {Campbell}},
  \bibinfo {author} {\bibfnamefont {M.}~\bibnamefont {Rambach}}, \bibinfo
  {author} {\bibfnamefont {K.}~\bibnamefont {Watanabe}}, \bibinfo {author}
  {\bibfnamefont {T.}~\bibnamefont {Taniguchi}},\ and\ \bibinfo {author}
  {\bibfnamefont {B.~D.}\ \bibnamefont {Gerardot}},\ }\bibfield  {title}
  {\bibinfo {title} {{Highly energy-tunable quantum light from
  moir{\'{e}}-trapped excitons}},\ }\href
  {https://doi.org/10.1126/sciadv.aba8526} {\bibfield  {journal} {\bibinfo
  {journal} {Science Advances}\ }\textbf {\bibinfo {volume} {6}},\ \bibinfo
  {pages} {eaba8526} (\bibinfo {year} {2020})}\BibitemShut {NoStop}%
\bibitem [{\citenamefont {Rosenberg}\ \emph {et~al.}(2018)\citenamefont
  {Rosenberg}, \citenamefont {Liran}, \citenamefont {Mazuz-Harpaz},
  \citenamefont {West}, \citenamefont {Pfeiffer},\ and\ \citenamefont
  {Rapaport}}]{Rosenberg2018}%
  \BibitemOpen
  \bibfield  {author} {\bibinfo {author} {\bibfnamefont {I.}~\bibnamefont
  {Rosenberg}}, \bibinfo {author} {\bibfnamefont {D.}~\bibnamefont {Liran}},
  \bibinfo {author} {\bibfnamefont {Y.}~\bibnamefont {Mazuz-Harpaz}}, \bibinfo
  {author} {\bibfnamefont {K.}~\bibnamefont {West}}, \bibinfo {author}
  {\bibfnamefont {L.}~\bibnamefont {Pfeiffer}},\ and\ \bibinfo {author}
  {\bibfnamefont {R.}~\bibnamefont {Rapaport}},\ }\bibfield  {title} {\bibinfo
  {title} {{Strongly interacting dipolar-polaritons}},\ }\href
  {https://doi.org/10.1126/sciadv.aat8880} {\bibfield  {journal} {\bibinfo
  {journal} {Science Advances}\ }\textbf {\bibinfo {volume} {4}},\ \bibinfo
  {pages} {eaat8880} (\bibinfo {year} {2018})}\BibitemShut {NoStop}%
\bibitem [{\citenamefont {Carusotto}\ and\ \citenamefont
  {Ciuti}(2013)}]{Carusotto2013}%
  \BibitemOpen
  \bibfield  {author} {\bibinfo {author} {\bibfnamefont {I.}~\bibnamefont
  {Carusotto}}\ and\ \bibinfo {author} {\bibfnamefont {C.}~\bibnamefont
  {Ciuti}},\ }\bibfield  {title} {\bibinfo {title} {{Quantum fluids of
  light}},\ }\href {https://doi.org/10.1103/RevModPhys.85.299} {\bibfield
  {journal} {\bibinfo  {journal} {Reviews of Modern Physics}\ }\textbf
  {\bibinfo {volume} {85}},\ \bibinfo {pages} {299} (\bibinfo {year} {2013})},\
  \Eprint {https://arxiv.org/abs/1205.6500} {arXiv:1205.6500} \BibitemShut
  {NoStop}%
\bibitem [{\citenamefont {Ołdziejewski}\ \emph {et~al.}(2021)\citenamefont
  {Ołdziejewski}, \citenamefont {Chiocchetta}, \citenamefont {Knörzer},\ and\
  \citenamefont {Schmidt}}]{Rafal2021}%
  \BibitemOpen
  \bibfield  {author} {\bibinfo {author} {\bibfnamefont {R.}~\bibnamefont
  {Ołdziejewski}}, \bibinfo {author} {\bibfnamefont {A.}~\bibnamefont
  {Chiocchetta}}, \bibinfo {author} {\bibfnamefont {J.}~\bibnamefont
  {Knörzer}},\ and\ \bibinfo {author} {\bibfnamefont {R.}~\bibnamefont
  {Schmidt}},\ }\href@noop {} {\bibinfo {title} {Excitonic tonks-girardeau and
  charge-density wave phases in monolayer semiconductors}} (\bibinfo {year}
  {2021}),\ \Eprint {https://arxiv.org/abs/2106.07290} {arXiv:2106.07290
  [cond-mat.mes-hall]} \BibitemShut {NoStop}%
\bibitem [{\citenamefont {Zomer}\ \emph {et~al.}(2014)\citenamefont {Zomer},
  \citenamefont {Guimar{\~{a}}es}, \citenamefont {Brant}, \citenamefont
  {Tombros},\ and\ \citenamefont {van Wees}}]{Zomer2014}%
  \BibitemOpen
  \bibfield  {author} {\bibinfo {author} {\bibfnamefont {P.~J.}\ \bibnamefont
  {Zomer}}, \bibinfo {author} {\bibfnamefont {M.~H.~D.}\ \bibnamefont
  {Guimar{\~{a}}es}}, \bibinfo {author} {\bibfnamefont {J.~C.}\ \bibnamefont
  {Brant}}, \bibinfo {author} {\bibfnamefont {N.}~\bibnamefont {Tombros}},\
  and\ \bibinfo {author} {\bibfnamefont {B.~J.}\ \bibnamefont {van Wees}},\
  }\bibfield  {title} {\bibinfo {title} {Fast pick up technique for high
  quality heterostructures of bilayer graphene and hexagonal boron nitride},\
  }\href {https://doi.org/10.1063/1.4886096} {\bibfield  {journal} {\bibinfo
  {journal} {Applied Physics Letters}\ }\textbf {\bibinfo {volume} {105}},\
  \bibinfo {pages} {013101} (\bibinfo {year} {2014})}\BibitemShut {NoStop}%
\bibitem [{\citenamefont {Telford}\ \emph {et~al.}(2018)\citenamefont
  {Telford}, \citenamefont {Benyamini}, \citenamefont {Rhodes}, \citenamefont
  {Wang}, \citenamefont {Jung}, \citenamefont {Zangiabadi}, \citenamefont
  {Watanabe}, \citenamefont {Taniguchi}, \citenamefont {Jia}, \citenamefont
  {Barmak}, \citenamefont {Pasupathy}, \citenamefont {Dean},\ and\
  \citenamefont {Hone}}]{Telford2018}%
  \BibitemOpen
  \bibfield  {author} {\bibinfo {author} {\bibfnamefont {E.~J.}\ \bibnamefont
  {Telford}}, \bibinfo {author} {\bibfnamefont {A.}~\bibnamefont {Benyamini}},
  \bibinfo {author} {\bibfnamefont {D.}~\bibnamefont {Rhodes}}, \bibinfo
  {author} {\bibfnamefont {D.}~\bibnamefont {Wang}}, \bibinfo {author}
  {\bibfnamefont {Y.}~\bibnamefont {Jung}}, \bibinfo {author} {\bibfnamefont
  {A.}~\bibnamefont {Zangiabadi}}, \bibinfo {author} {\bibfnamefont
  {K.}~\bibnamefont {Watanabe}}, \bibinfo {author} {\bibfnamefont
  {T.}~\bibnamefont {Taniguchi}}, \bibinfo {author} {\bibfnamefont
  {S.}~\bibnamefont {Jia}}, \bibinfo {author} {\bibfnamefont {K.}~\bibnamefont
  {Barmak}}, \bibinfo {author} {\bibfnamefont {A.~N.}\ \bibnamefont
  {Pasupathy}}, \bibinfo {author} {\bibfnamefont {C.~R.}\ \bibnamefont
  {Dean}},\ and\ \bibinfo {author} {\bibfnamefont {J.}~\bibnamefont {Hone}},\
  }\bibfield  {title} {\bibinfo {title} {{Via Method for Lithography Free
  Contact and Preservation of 2D Materials}},\ }\href
  {https://doi.org/10.1021/acs.nanolett.7b05161} {\bibfield  {journal}
  {\bibinfo  {journal} {Nano Letters}\ }\textbf {\bibinfo {volume} {18}},\
  \bibinfo {pages} {1416} (\bibinfo {year} {2018})}\BibitemShut {NoStop}%
\bibitem [{\citenamefont {Jung}\ \emph {et~al.}(2019)\citenamefont {Jung},
  \citenamefont {Choi}, \citenamefont {Nipane}, \citenamefont {Borah},
  \citenamefont {Kim}, \citenamefont {Zangiabadi}, \citenamefont {Taniguchi},
  \citenamefont {Watanabe}, \citenamefont {Yoo}, \citenamefont {Hone},\ and\
  \citenamefont {Teherani}}]{Jung2019}%
  \BibitemOpen
  \bibfield  {author} {\bibinfo {author} {\bibfnamefont {Y.}~\bibnamefont
  {Jung}}, \bibinfo {author} {\bibfnamefont {M.~S.}\ \bibnamefont {Choi}},
  \bibinfo {author} {\bibfnamefont {A.}~\bibnamefont {Nipane}}, \bibinfo
  {author} {\bibfnamefont {A.}~\bibnamefont {Borah}}, \bibinfo {author}
  {\bibfnamefont {B.}~\bibnamefont {Kim}}, \bibinfo {author} {\bibfnamefont
  {A.}~\bibnamefont {Zangiabadi}}, \bibinfo {author} {\bibfnamefont
  {T.}~\bibnamefont {Taniguchi}}, \bibinfo {author} {\bibfnamefont
  {K.}~\bibnamefont {Watanabe}}, \bibinfo {author} {\bibfnamefont {W.~J.}\
  \bibnamefont {Yoo}}, \bibinfo {author} {\bibfnamefont {J.}~\bibnamefont
  {Hone}},\ and\ \bibinfo {author} {\bibfnamefont {J.~T.}\ \bibnamefont
  {Teherani}},\ }\bibfield  {title} {\bibinfo {title} {{Transferred via
  contacts as a platform for ideal two-dimensional transistors}},\ }\href
  {https://doi.org/10.1038/s41928-019-0245-y} {\bibfield  {journal} {\bibinfo
  {journal} {Nature Electronics}\ }\textbf {\bibinfo {volume} {2}},\ \bibinfo
  {pages} {187} (\bibinfo {year} {2019})}\BibitemShut {NoStop}%
\bibitem [{\citenamefont {Wilson}\ \emph {et~al.}(2017)\citenamefont {Wilson},
  \citenamefont {Nguyen}, \citenamefont {Seyler}, \citenamefont {Rivera},
  \citenamefont {Marsden}, \citenamefont {Laker}, \citenamefont
  {Constantinescu}, \citenamefont {Kandyba}, \citenamefont {Barinov},
  \citenamefont {Hine}, \citenamefont {Xu},\ and\ \citenamefont
  {Cobden}}]{Wilson2017}%
  \BibitemOpen
  \bibfield  {author} {\bibinfo {author} {\bibfnamefont {N.~R.}\ \bibnamefont
  {Wilson}}, \bibinfo {author} {\bibfnamefont {P.~V.}\ \bibnamefont {Nguyen}},
  \bibinfo {author} {\bibfnamefont {K.}~\bibnamefont {Seyler}}, \bibinfo
  {author} {\bibfnamefont {P.}~\bibnamefont {Rivera}}, \bibinfo {author}
  {\bibfnamefont {A.~J.}\ \bibnamefont {Marsden}}, \bibinfo {author}
  {\bibfnamefont {Z.~P.}\ \bibnamefont {Laker}}, \bibinfo {author}
  {\bibfnamefont {G.~C.}\ \bibnamefont {Constantinescu}}, \bibinfo {author}
  {\bibfnamefont {V.}~\bibnamefont {Kandyba}}, \bibinfo {author} {\bibfnamefont
  {A.}~\bibnamefont {Barinov}}, \bibinfo {author} {\bibfnamefont {N.~D.}\
  \bibnamefont {Hine}}, \bibinfo {author} {\bibfnamefont {X.}~\bibnamefont
  {Xu}},\ and\ \bibinfo {author} {\bibfnamefont {D.~H.}\ \bibnamefont
  {Cobden}},\ }\bibfield  {title} {\bibinfo {title} {Determination of band
  offsets, hybridization, and exciton binding in 2d semiconductor
  heterostructures},\ }\href {https://doi.org/10.1126/sciadv.1601832}
  {\bibfield  {journal} {\bibinfo  {journal} {Science Advances}\ }\textbf
  {\bibinfo {volume} {3}},\ \bibinfo {pages} {e1601832} (\bibinfo {year}
  {2017})}\BibitemShut {NoStop}%
\bibitem [{\citenamefont {Laturia}\ \emph {et~al.}(2018)\citenamefont
  {Laturia}, \citenamefont {de~Put},\ and\ \citenamefont
  {Vandenberghe}}]{Laturia2018}%
  \BibitemOpen
  \bibfield  {author} {\bibinfo {author} {\bibfnamefont {A.}~\bibnamefont
  {Laturia}}, \bibinfo {author} {\bibfnamefont {M.~L.~V.}\ \bibnamefont
  {de~Put}},\ and\ \bibinfo {author} {\bibfnamefont {W.~G.}\ \bibnamefont
  {Vandenberghe}},\ }\bibfield  {title} {\bibinfo {title} {Dielectric
  properties of hexagonal boron nitride and transition metal dichalcogenides:
  from monolayer to bulk},\ }\bibfield  {journal} {\bibinfo  {journal} {npj 2D
  Materials and Applications}\ }\textbf {\bibinfo {volume} {2}},\ \href
  {https://doi.org/10.1038/s41699-018-0050-x} {10.1038/s41699-018-0050-x}
  (\bibinfo {year} {2018})\BibitemShut {NoStop}%
\bibitem [{\citenamefont {Smoleński}\ \emph {et~al.}(2019)\citenamefont
  {Smoleński}, \citenamefont {Cotlet}, \citenamefont {Popert}, \citenamefont
  {Back}, \citenamefont {Shimazaki}, \citenamefont {Knüppel}, \citenamefont
  {Dietler}, \citenamefont {Taniguchi}, \citenamefont {Watanabe}, \citenamefont
  {Kroner},\ and\ \citenamefont {Imamoglu}}]{Smolenski2018}%
  \BibitemOpen
  \bibfield  {author} {\bibinfo {author} {\bibfnamefont {T.}~\bibnamefont
  {Smoleński}}, \bibinfo {author} {\bibfnamefont {O.}~\bibnamefont {Cotlet}},
  \bibinfo {author} {\bibfnamefont {A.}~\bibnamefont {Popert}}, \bibinfo
  {author} {\bibfnamefont {P.}~\bibnamefont {Back}}, \bibinfo {author}
  {\bibfnamefont {Y.}~\bibnamefont {Shimazaki}}, \bibinfo {author}
  {\bibfnamefont {P.}~\bibnamefont {Knüppel}}, \bibinfo {author}
  {\bibfnamefont {N.}~\bibnamefont {Dietler}}, \bibinfo {author} {\bibfnamefont
  {T.}~\bibnamefont {Taniguchi}}, \bibinfo {author} {\bibfnamefont
  {K.}~\bibnamefont {Watanabe}}, \bibinfo {author} {\bibfnamefont
  {M.}~\bibnamefont {Kroner}},\ and\ \bibinfo {author} {\bibfnamefont
  {A.}~\bibnamefont {Imamoglu}},\ }\bibfield  {title} {\bibinfo {title}
  {Interaction-induced shubnikov-de haas oscillations in optical conductivity
  of monolayer {MoSe$_2$}},\ }\href
  {https://doi.org/10.1103/PhysRevLett.123.097403} {\bibfield  {journal}
  {\bibinfo  {journal} {Physical Review Letters}\ }\textbf {\bibinfo {volume}
  {123}},\ \bibinfo {pages} {097403} (\bibinfo {year} {2019})},\ \Eprint
  {https://arxiv.org/abs/1812.08772} {1812.08772} \BibitemShut {NoStop}%
\end{thebibliography}%

% \newpage
%\clearpage

	%In-plane electric fields can be generated along the edge of the TG by applying a voltage bias between TG and BG (dashed black lines).  
	%Applying a voltage bias between the top and bottom gates leads to inhomogeneous in-plane electric fields (dashed black lines) along the TG edge. Strong fields can be generated in the p-i-n regime where the monolayer is electron doped (red) and the region underneath the top gate is hole doped (blue). 

%\section*{Extended Figures}
\renewcommand{\figurename}{Extended Fig.}
\setcounter{figure}{0}

%\clearpage
%\newpage

\begin{center}
\begin{figure*}[ht!]
	\includegraphics[width=11cm]{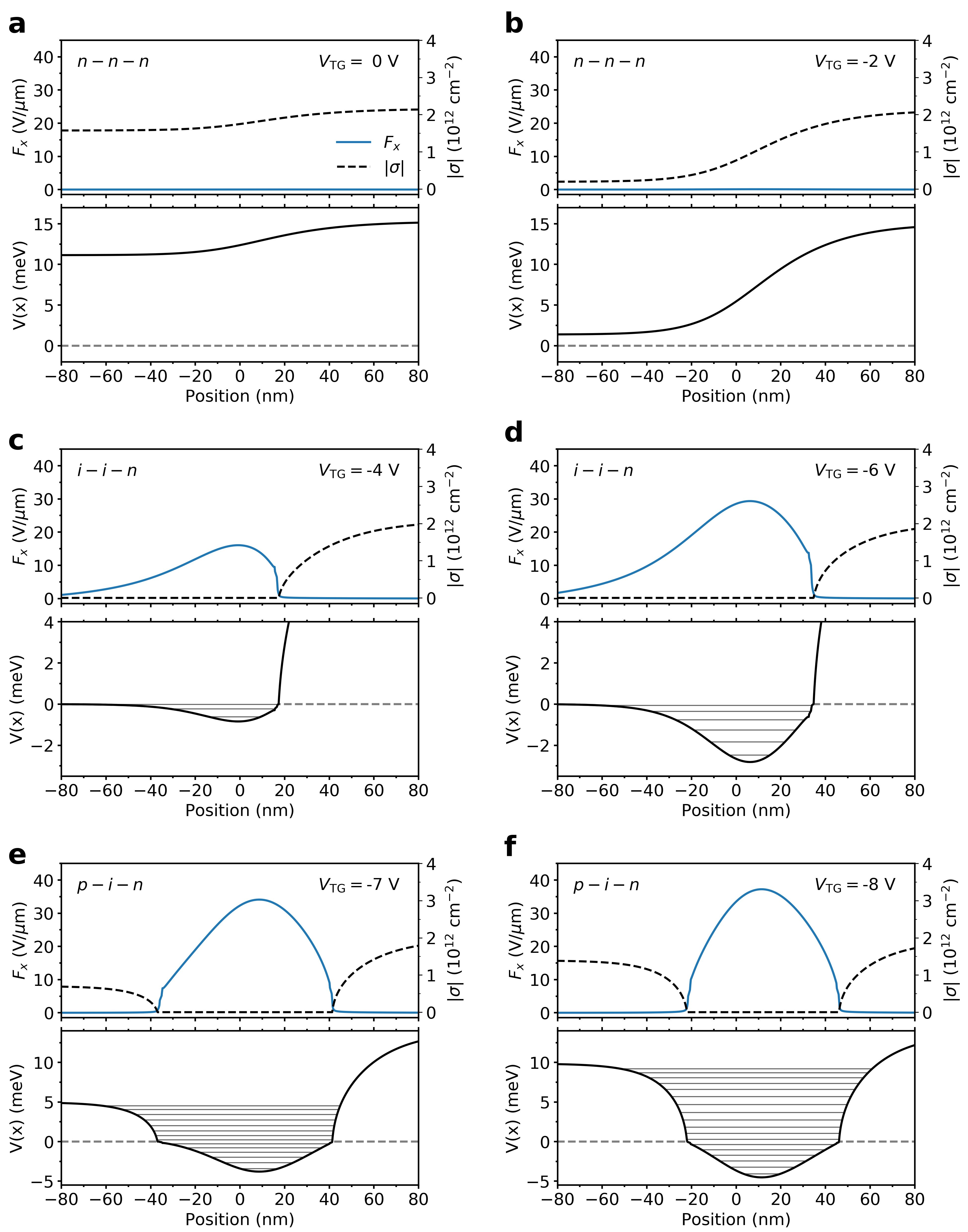}
	\caption{\textbf{Electrostatic simulations of the device.} Magnitude of charge density $|\sigma(x)|$, in-plane electric field $|F_x|$ and exciton confining potential $V(x)$ as a function of position for varying top gate voltages $\vtg$. The bottom gate extends over the entire plotted range of the position from $-80$\,nm to $80$\,nm. The top gate extends from $-80$\,nm to $0$\,nm, with the edge at $x = 0$. The different charging configurations, namely n-n-n (\bfA, \bfB), i-i-n (\bfC,\bfD) and p-i-n (\bfE,\bfF) show the evolution of the confinement potential as a function of $\vtg$. }  
	\label{fig:Fx_nc}
\end{figure*} 
\end{center}

\begin{figure*}[ht!]
	\includegraphics[width=7cm]{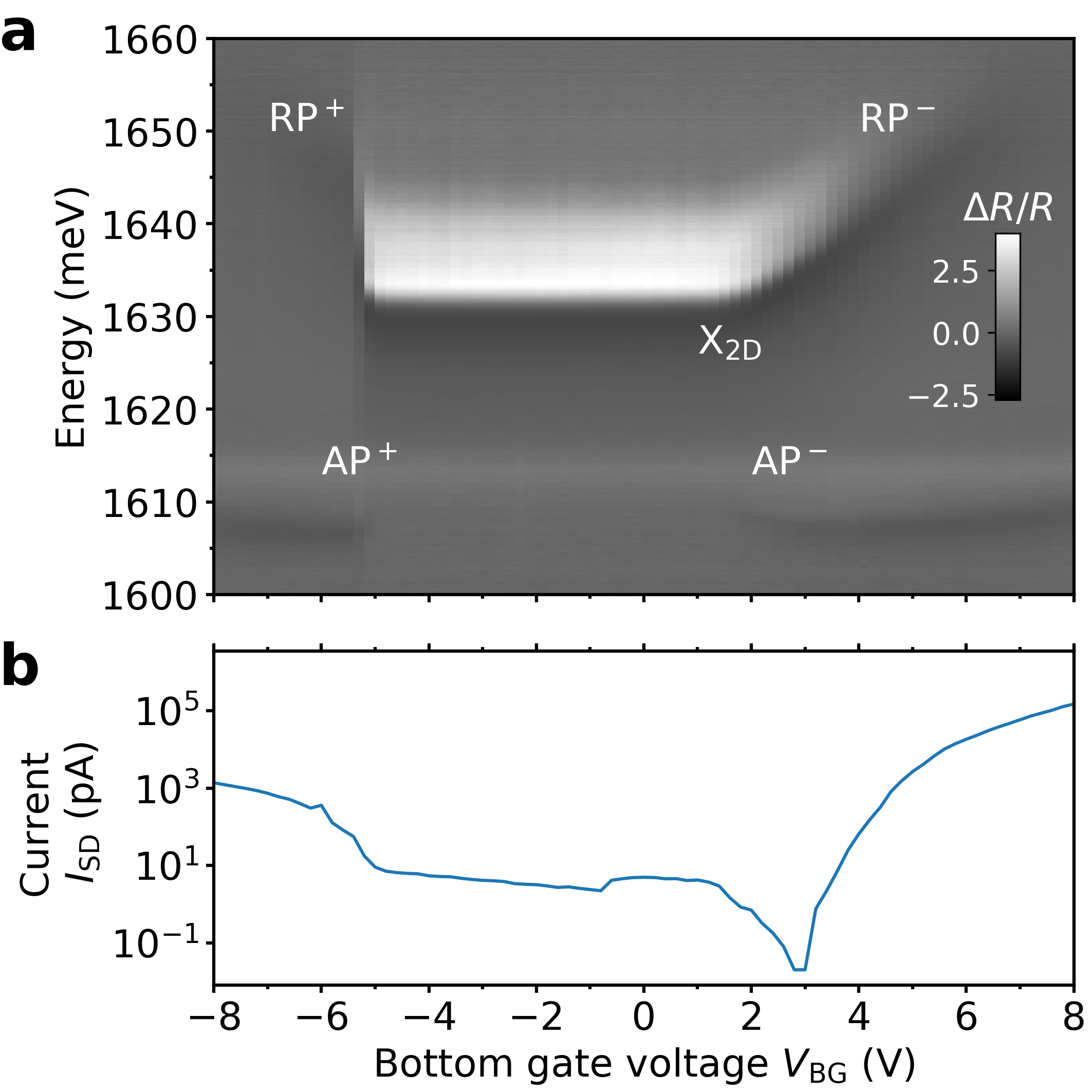}
	\caption{\textbf{Characterization of Device 1.} (\bfA) Representative reflectance measurement as a function of $\vbg$ performed in region I, away from the TG. We observe the neutral exciton state from $-5.5\,\mathrm{V} < \vbg < 2\,\mathrm{V}$, flanked by the repulsive polaron on the electron side (RP$^-$: $\vbg > 2\,\mathrm{V}$) and the hole side (RP$^+$: $\vbg < -5.5\,\mathrm{V}$). In addition, we observe the corresponding electron and hole side attractive polaron branches (AP$^-$ and AP$^+$). The exciton energy $E_\mathrm{X,2D}$ shows spatial variation of $\approx 2\,$meV across the sample due to disorder and strain. (\bfB) Source-Drain transport measurements through the monolayer MoSe$_2$, with constant source-drain bias $V_\mathrm{SD} = 2\,\mathrm{V}$ as a function of $\vbg$. We observe that the onset of doping coincides with the onset of the polaron branches in (\bfA), demonstrating that the optical spectroscopic signatures are reliable probes of the doping configuration in our device. }  
	\label{fig:Dev1}
\end{figure*}

	%Optical micrograph of Device 1 studied in this work. The dashed white line corresponds to the outline of the monolayer MoSe$_2$ flake. The inset shows an AFM scan of the split gate (inset scale bar: $200\,$nm). (\bfB) Typical broadband reflectance spectra of the encapsulated TMD monolayer as a function of $\vbg$, taken at a position away from the TG (marked by $*$ in \bfA). This shows the exciton resonance in the charge neutral regime and the attractive and repulsive polaron branches that emerge in the electron and hole doped regimes. (\bfC) Applying a negative $\vtg$ depletes the region below the TG leaving only a narrow 1D channel for electron transport, which is effectively pinched off at large negative $\vtg$. Source-drain measurements as a function of $\vtg$ for different $\vbg$ show a nonlinear step-like rise in current, indicating confinement of electrons in the 1D channel.
\newpage

\begin{figure*}[ht!]
	\includegraphics{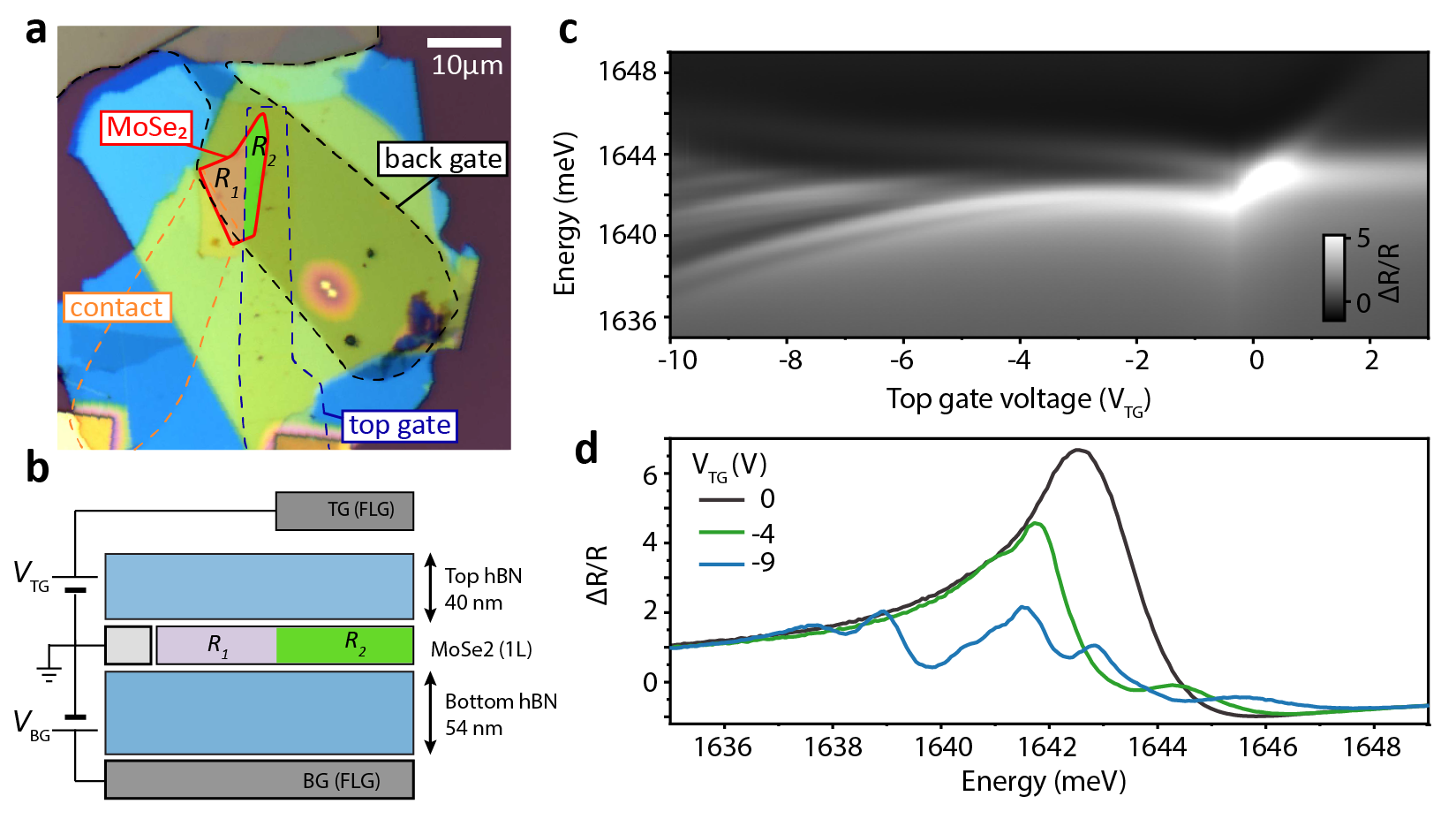}
	\caption{\textbf{Electrically tunable quantum confinement in Device 2.} (\bfA) Optical micrograph of Device 2, where the outline of the MoSe$_2$ monolayer is indicated by the red line. The top and bottom gates, made of few-layer graphene (FLG), are indicated by dashed black lines. (\bfB) Schematic diagram of Device 2. (\bfC) Spectra taken at the edge of the FLG top gate at $\vbg=1$\,V shows the emergence of discrete states below the 2D continuum, in excellent agreement with the observations in Device 1 (Fig.\,\ref{fig:WL}\,\bfB).(\bfD) Spectral linecuts at $\vtg = 0\,$V, $-4\,$V, and $-9\,$V show the emergence of confined states as the potential is made deeper. The exact voltage range and the magnitude of red shift may differ between the two devices due to design differences, in particular the thickness of h-BN spacers.}  
	\label{fig:Dev2}
\end{figure*}

\newpage

\begin{figure*}
	\includegraphics[width=8cm]{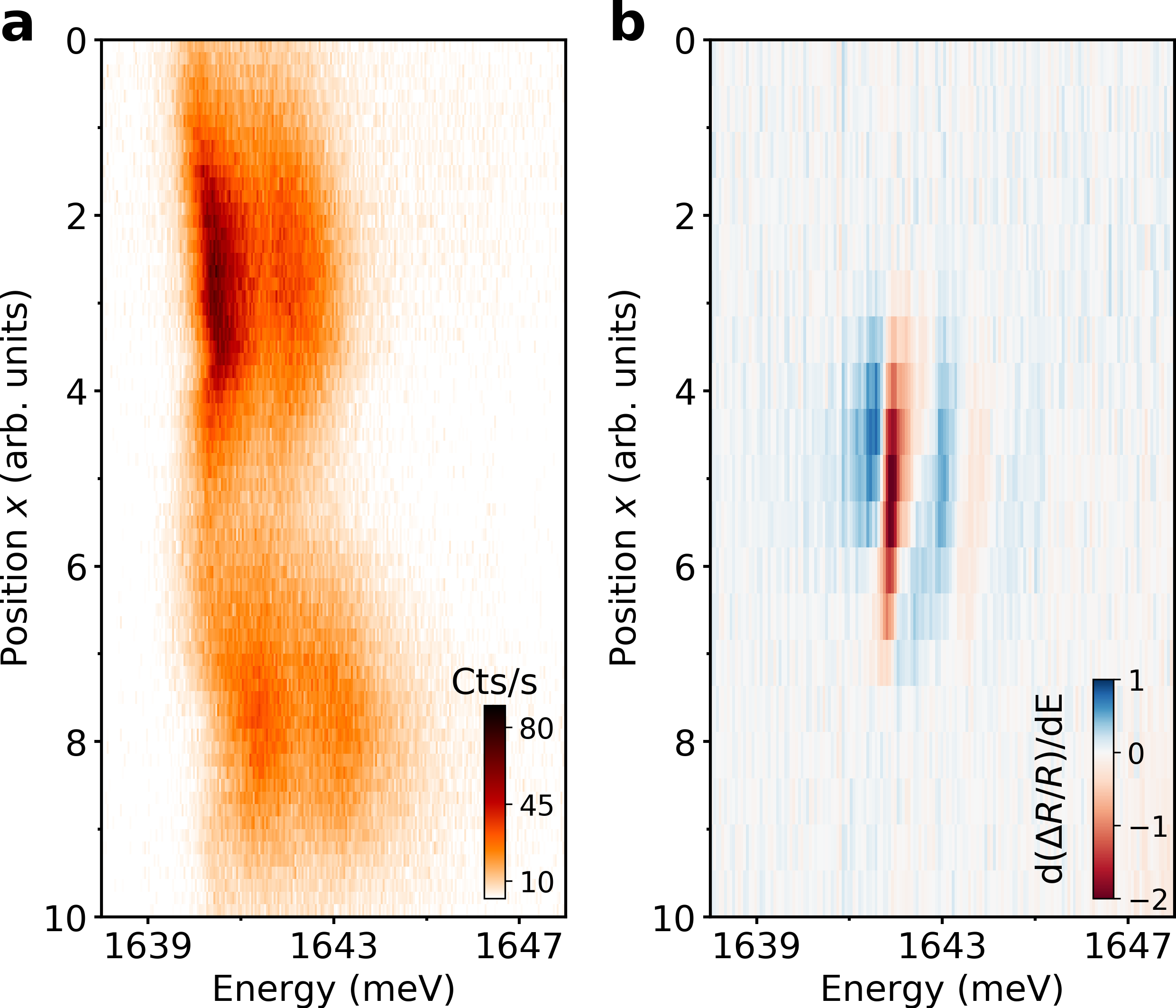}
	\caption{\textbf{Exciton confinement along the TG edge.} We measure the optical response of confined excitons as the position of the optical spot is scanned across the TG edge using $x-y$ nanopositioners. (\bfA) Position dependent PL spectra in Device 1. The TG in this case consists of two edges separated by $\approx 1\,\mu$m. Hence, the discrete confined states appear at two spatial locations and almost vanish in the intermediate region. (\bfB) Position-dependent reflectance spectra in Device 2, which features a single TG edge. In both cases, resolution is determined by the diffraction-limited spot size.}  
	\label{fig:1Dedge}
\end{figure*}

% \begin{figure*}
% 	\includegraphics[width=8cm]{Figures/Ext_PLWL.png}
% 	\caption{\textbf{Comparison of reflection and PL spectra.} (\bfA) The reflection spectra shown in Fig.\,\ref{fig:WL} {\bfB} of the main text, which is measured at low white light power ($\sim 10\,$nW). (\bfB) $\vtg$-dependent PL spectra of confined excitons taken at the same position, with PL laser power $\sim 20\,\mu$W. (\bfC) Reflection spectra measured with simultaneous PL and white light illumination shows confined states whose voltage dependence is consisted with the PL lines shown in (\bfB). This indicates that the difference between reflection (\bfA) and PL (\bfB) measurements arises mainly due to the higher excitation power of the latter, which modifies the charge distribution in the p- and n-doped regions and the corresponding electric field distribution in the i-region. This is further supported by the fact that when the PL illumination is present, the extent of the charge neutral regime is reduced, and the onset of RP$^+$ branch occurs already at $\vtg \sim -4.5\,$V rather than $\vtg = -6\,$V as seen in (\bfA).    }  
% 	\label{fig:PLWL}
% \end{figure*} 

\clearpage
%\newpage

\section{Supplementary Information}
\renewcommand{\figurename}{SI Fig.}
\setcounter{figure}{0}
\renewcommand{\thefigure}{S\arabic{figure}}

% % \startcontents[sections]
% % \printcontents[sections]{}{1}{}

\subsection{Sample fabrication}

All MoSe$_2$, h-BN and graphene flakes used to assemble the devices presented in the main text above are obtained through mechanical exfoliation of bulk crystals. The flakes are stacked into a heterostructure in an inert Ar atmosphere inside a glovebox using a standard dry polymer transfer technique \cite{Zomer2014}. For Device 1 the MoSe$_2$ monolayer is encapsulated using $\sim 30$\,nm thick h-BN and deposited on a pre-patterned Ti/Au ($3$\,nm/$10$\,nm) bottom gate. Pd/Au ($20$\,nm/$30$\,nm) contacts to the MoSe$_2$ layer are embedded in the top h-BN layer and prepared using the \textit{Via}--contacting method \cite{Telford2018,Jung2019}. Subsequently, a $200$\,nm wide split top gate electrode featuring a $100$\,nm gap is formed by evaporating Ti/Au ($3$\,nm/ $10$\,nm). This renders the top gate optically transparent and therefore allows to probe the optical properties of the monolayer underneath while the local charge density is being altered. Metal electrodes to all contacts and both gates are formed with Ti/Au ($5$\,nm/ $85$\,nm). Device 2 has a similar architecture as Device 1. However, all gate electrodes and the contact to the MoSe$_2$ layer were made using few-layer graphene. Additionally, the top/ bottom h-BN spacer layers were chosen to be thicker ($40$\,nm and $54$\,nm, respectively).

\subsection{Experimental setup}

% \begin{figure}[ht!]
% 	\includegraphics[width=7.5cm]{Figures/FigS2.png}
% 	\caption{\textbf{Simultaneous measurement of optical reflectance and source-drain current.} The onset of source-drain current approximately coincides with the onset of blue shift of the repulsive polaron resonances.}  
% 	\label{fig:WL_Isd}
% \end{figure}

We perform our optical experiments in a confocal microscope setup. The sample is mounted on $x$-$y$-$z$ piezo-electric stages located inside a stainless steel tube, which is immersed in a liquid helium bath cryostat. The steel tube is filled with 20 mbar helium exchange gas to maintain a sample temperature of $\sim 4.2$ K. Free-space optical access to the sample is enabled through a glass window on top of the tube. White light (WL) reflectance and photoluminescence is measured using a broadband light-emitting diode centered at $760\,$nm and a tunable single-mode Ti:Sapphire laser as the excitation source, respectively. The light from the source is focused to a diffraction-limited spot through a high numerical aperture lens (NA = 0.68). The light reflected / emitted from the sample is then collected using the same lens, separated from the incident light by a beam splitter, coupled into a single-mode fiber and imaged on a spectrometer equipped with a liquid-nitrogen-cooled charge-coupled-device (CCD). For WL measurements an excitation power of a few tens of nW, and for PL a few $\mu$W is maintained. The polarization-resolved PL measurements are carried out with an angle-scanning polarizer placed in the emission path.

\subsection{Electrostatic simulation of device geometry}

In this section, we discuss the finite-element calculations of our devices, which provide us quantitative information of the in-plane fields, charge densities and the corresponding confinement potentials. These computations are performed using the \emph{Electrostatics} package in COMSOL. For all our simulations, we assume temperature $T=0\,$K. We employ the Thomas--Fermi approximation and model the MoSe$_2$ monolayer as a single sheet of charge with density,
\begin{align}
    \sigma(x) = \sigma_n(x) + \sigma_p(x)
\end{align}
where $\sigma_n$ and $\sigma_p$ are the electron and hole charge densities, which are in turn given by,
\begin{align}
    \sigma_n(x) &= -e \int_{E_C(V(x))}^{E_F} \mathcal{D}(E) dE, &E_F > E_C(V(x)) \nonumber \\
    &= -e \mathcal{D}(E)(E_F-E_C), \\
    \sigma_p(x) &= e \int_{E_F}^{E_V(V(x))} \mathcal{D}(E) dE, &E_F < E_V(V(x)) \nonumber \\
    &= e \mathcal{D}(E)(E_V-E_F).
\end{align}
Here, $\mathcal{D}(E) = \frac{g_s g_v m^*}{2\pi\hbar^2}$ is the 2D density of states for electrons and holes in the semiconductor, where $g_s = 1$ is the spin degeneracy and $g_v = 2$ is the valley degeneracy. $E_C$ and $E_V$ are the conduction and valence band edge energies, which depend on the local electrostatic potential. $E_\mathrm{F}$ is the Fermi energy, which is set by the alignment of the contact work function with respect to the band edges, and is assumed to be constant across the device.

The simulated device geometry is depicted in Fig.\,1\,{\bfA} of the main text. The sheet of charge is encapsulated by $30$\,nm thick h-BN slabs and contacted by ohmic electrodes. Furthermore, we include two gates with partial overlap. The bottom gate is kept at a fixed bias of $4$\,V and ensures a global electron doping throughout the semiconductor. The voltage on the top gate is varied from $0$\,V to $-10$\,V in our simulations. The material parameters assumed for this calculation are the following: MoSe$_2$ bandgap $E_b = 1.85$\,eV, Fermi level offset relative to the valence band edge at zero potential $E_F-E_V(V=0)=0.99$\,eV \cite{Wilson2017}, electron effective mass $m_{n}^* = 0.7\,m_e$ \cite{Larentis2018}, hole effective mass $m_{p}^* = 0.6\,m_e$ \cite{Zhang2014,Goryca2019}, out-of-plane dielectric constant  $\varepsilon_{\perp} = 3.76$, and in-plane dielectric constant $\varepsilon_{\parallel} = 6.93$ for h-BN \cite{Laturia2018}.

In this manner we obtain the spatial charge density and in-plane electric field distribution for varying top gate voltages $\vtg$, which are depicted in the top panel of each subplot in Fig.\,\ref{fig:Fx_nc}. The bottom gate extends over the entire plotted range of the position from $-80$\,nm to $80$\,nm, whereas the top gate only ranges from $-80$\,nm to $0$\,nm, with the edge located at $x=0$. We can identify three distinct regimes as we vary $\vtg$, which we identify according to the doping state in regions I, II and III. For $\vtg > -4\,$V, we are in the n-n-n regime, where all three regions are n-doped, but the electron density varies spatially, as shown in Fig.\,\ref{fig:Fx_nc}\,{\bfA} and {\bfB}. As we decrease $\vtg$ further, we deplete regions II and III completely and hence we are in the i-i-n regime, which is accompanied by a large increase in magnitude of the in-plane electric field $|F_x|$ (Fig.\,\ref{fig:Fx_nc}\,{\bfC} and {\bfD}). While the maximum of the field distribution is located close to the top gate edge, owing to the large lateral extent of the neutral region, the in-plane field persists even under the TG and exhibits a spatial asymmetry. The i-i-n regime persists until the onset of hole-doping in region II, which occurs in our simulations (and experiments) at $\vtg < -6\,$V (see Fig.\,\ref{fig:Fx_nc}\,{\bfE} and {\bfF}). As hole-doping starts, only a narrow $\sim 60$\,nm wide neutral region remains, located at the edge of the TG and flanked by a steep increase in charge density. This is the p-i-n regime. As a consequence, the in-plane field distribution also becomes concentrated in this narrow region due to screening in the neighbouring charged areas. Lowering $\vtg$ further pushes the neutral junction region further away from the TG, thereby making the electric field distribution increasingly symmetric. Ultimately, at $\vtg = -10$\,V we obtain a sizeable in-plane electric field, with a maximum of $|F_x| \sim 40$\,V/$\mu$m.

From these quantities we determine the total excitonic confining potential as outlined in Eq.\,(1) of the main text. The dc Stark shift contribution is computed by assuming an exciton polarizability $\alpha = 6.5\,\mathrm{eV\,nm}^2/\mathrm{V}^2$ \cite{Cavalcante2018}. The repulsive polaron shift is determined by empirically extracting an effective exciton-electron coupling strength $\beta \simeq 0.7\,\mu\mathrm{eV}\mu\mathrm{m^2}$ from our experimental data. It corresponds to the slope of a linear function fitted to the density-dependent blue shift of the repulsive polaron in the reflectance data, shown in Fig.\,2\,{\bfB} of the main text.

The resulting potential experienced by the exciton in its center-of-mass frame for various $\vtg$ is depicted in the lower panel of each subplot in Extended Fig.\,\ref{fig:Fx_nc}. Additionally, whenever appropriate, we also show the numerically calculated discrete eigenstates associated with the confining potential. A potential well starts to form only at $\vtg = -4$\,V, but is not strong enough to lead to discernible quantization (Fig.\,\ref{fig:Fx_nc}\,{\bfC}). However, the 2D excitonic state may exhibit a small red shift. At $\vtg < -4$\,V the Stark shift contribution starts to become significant and leads to the formation of a much narrower potential well localized in close proximity to the TG edge.  Here, in the i-i-n regime, the continuum is given by the 2D free exciton energy $E_\mathrm{X,2D}$ and the confinement is solely driven by the dc Stark shift. As the gate voltage is lowered further to $\vtg < -6$\,V, we enter the p-i-n regime (Fig.\,\ref{fig:Fx_nc}\,{\bfE} an {\bfF}), where the continuum is solely determined by the hole or electron RP energy, depending on which charge has lower density. This leads to the striking observation of quantized states above the free exciton energy $E_\mathrm{X,2D}$.

\subsection{Doping properties}

\begin{figure*}[ht]
	\includegraphics[width=12cm]{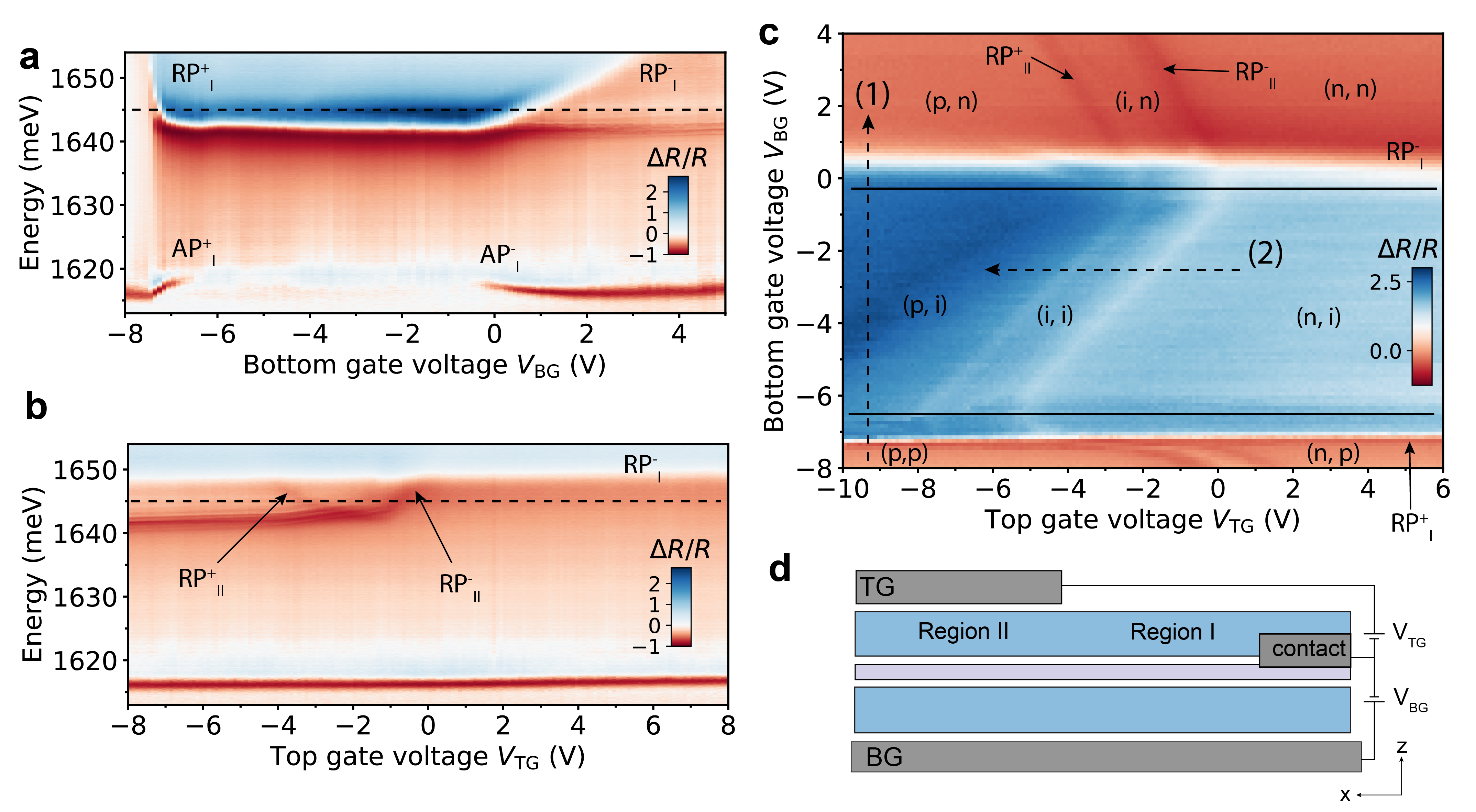}
	\caption{\textbf{Doping characteristics tracked using repulsive polaron resonances.} Since the RP states exist in regions I and II of the device and on electron and hole sides, we use the notation RP$^\mathrm{charge}_\mathrm{region}$ to denote the different states. (\bfA) Normalized reflectance $\Delta R/R$ measured for fixed $\vtg = -6\,$V as a function of $\vbg$, where we mainly observe features from region I. (\bfB) $\Delta R/R$ measured for fixed $\vbg = 1.8\,$V as a function of $\vtg$, where we observe the spectral changes in Region II. In addition to the features from regions II, we see the repulsive polaron resonance from region I at $E \approx 1648\,$meV which is not affected by $\vtg$.	(\bfC) Normalized reflectance taken at a fixed energy $E_{X0} + \Gamma/2 = 1645\,$meV, where $\Gamma$ is the bare 2D exciton linewidth. The horizontal solid lines demarcate the n-doped, neutral and p-doped regimes in region I. The doping configurations in regions I and II in different voltage regimes are identified. For example, in the top left corner, which is the relevant regime for this work, the doping configuration is $(\mathrm{II, I}) \equiv (\mathrm{p,n})$). (\bfD) Schematic of the device indicating regions I and II, as well as the position of the electrical contact to the monolayer.}  
	\label{fig:doping}
\end{figure*}

To understand the doping behaviour in our devices, we consider three spatial regions as follows: (I) the region away from the TG that is affected only by the BG, (II) the region directly underneath the TG, and (III) the narrow region between I and II. Due to the diffraction-limited spot size, we simultaneously measure the combined optical response of all three regions. We note that the energy of the repulsive polaron state is a sensitive probe of the charge density. Therefore, the charging behavior of the p-i-n junction in our devices can be understood by measuring the reflectance spectra as a function of $\vbg$ and $\vtg$. Specifically, measuring the reflectance at fixed energy $E = E_\mathrm{X,2D} + \Gamma/2 =  1645\,$meV (dashed line in Fig.\,\ref{fig:doping}\,\bfA and \bfB) allows to identify the doping configuration.

In Fig.\,\ref{fig:doping}\,\bfA, we show the reflectance spectra as a function of $\vbg$ (for fixed $\vtg$), which shows the typical charging behavior that exhibits neutral exciton, RP and AP branches on the n- and p-doping regimes in region I. Moving from right to left along the horizontal dashed line takes us from the n- to i- to p-doped regimes in region I. Similarly, in Fig.\,\ref{fig:doping}\,\bfB, we show the reflectance as a function of $\vtg$ (for fixed $\vbg$), which also exhibits the neutral and AP branches. Once again, moving from right to left along the dashed line in {\bfB} takes us from n- to i- to p- doped regimes in region II. We note that, in both cases, as we move along the dashed lines, we observe two RP states: RP$^-$ and $RP^+$. These resonances demarcate regions of electron and hole doping. On the right of the RP$^-$, we are always electron doped, whereas on the left of RP$^+$, we are always hole doped. In between, we are in the neutral regime. 

Using this method, we can monitor the charge configurations in both regions I and II, by measuring the reflectance at fixed energy as a function of $\vtg$ and $\vbg$, as shown in Fig.\,\ref{fig:doping}\,\bfC. Due to the mechanism explained above, we observe that the RP$^+$ and RP$^-$ resonance exhibit a zig-zag shape in the $\vbg-\vtg$ plane. Since the energy of the RP state is proportional to charge density, moving on the zig-zag line amounts to moving along a path of constant charge density. To identify the charging configurations, we use the notation (II,I). For example, $(p,n)$ refers to p-doping in II and n-doping in I. In general, moving vertically at fixed $\vtg$ tunes the doping in region I. For example, moving along path 1 takes the doping configuration from (p,p) to (p,i) to (p,n). Similarly, moving horizontally keeping $\vbg$ fixed tunes the doping in region II. 

We observe in Fig.\,\ref{fig:doping}\,{\bfC}\, that all charging configurations are possible in our device. This is surprising considering that the electrical contact to the TMD monolayer exists only in region I and not in region II. Therefore, any charges that have to reach region II from the contacts (in region I) have to necessarily pass through region I in an equilibrium setting. For example, in the top left corner we observe the p-n regime that is relevant for this work. In this voltage range, it is energetically unfavorable for holes to reach region II by traversing region I which is n-doped. To explain this, we consider a charge injection mechanism based on optical doping, which is a local effect.

When region I is neutral, a voltage bias between TG and BG results in electric fields ($F \propto \vbg - \vtg$) which can lead to a trapping potential for charges under the TG. However, since the Fermi energy is in the gap, in thermodynamic equilibrium it is not energetically favourable for electrons or holes to populate this trap. Nevertheless, upon optical excitation, excitons can dissociate under the strong in-plane electric field generated at the edges of the TG. Thus, individual charge carriers are left behind which are accelerated towards region II, where they ultimately get trapped. On the other hand, the generated field between TG and BG also leads to a finite rate of charges tunnelling out of the TMD in the out-of-plane direction. We suspect that this interplay between charge injection through exciton dissociation and charge removal by means of leakage currents defines the doping configuration in our device. The impact of both mechanisms is influenced by the electric field magnitude. Correspondingly, the observed charge density in region II also scales linearly $\sigma \propto \vbg - \vtg$. The doping situation is different when the device is globally n- or p-doped. In order to reach charge neutrality, $\vtg$ now needs to be tuned to a voltage that removes the charges under the top gate, therefore at the equal density line $\sigma \propto \vtg + \vbg$. We expect that in this scenario the doping of opposite charge carriers in the region II also arises from the aforementioned optical doping mechanism.

% % This optical doping mechanism provides a reliable means of doping the region under the TG. In the global charge neutrality region, the charge density in region II depends on the in-plane electric field which induces dissociation and therefore scales linearly with $\sigma \propto \vbg - \vtg$. On the other hand, the doping situation is different when the device is globally $n$ or $p$ doped. In order to reach charge neutrality, $V_{TG}$ now needs to be tuned to a voltage that removes the charges under the top gate, therefore the equal density line follows $V_{TG} + V_{BG} = \mathrm{constant}$. We expect the doping of opposite charge carriers in the region II to arise from the same optical doping mechanism described above.

\begin{figure}[ht!]
	\includegraphics[width=7.5cm]{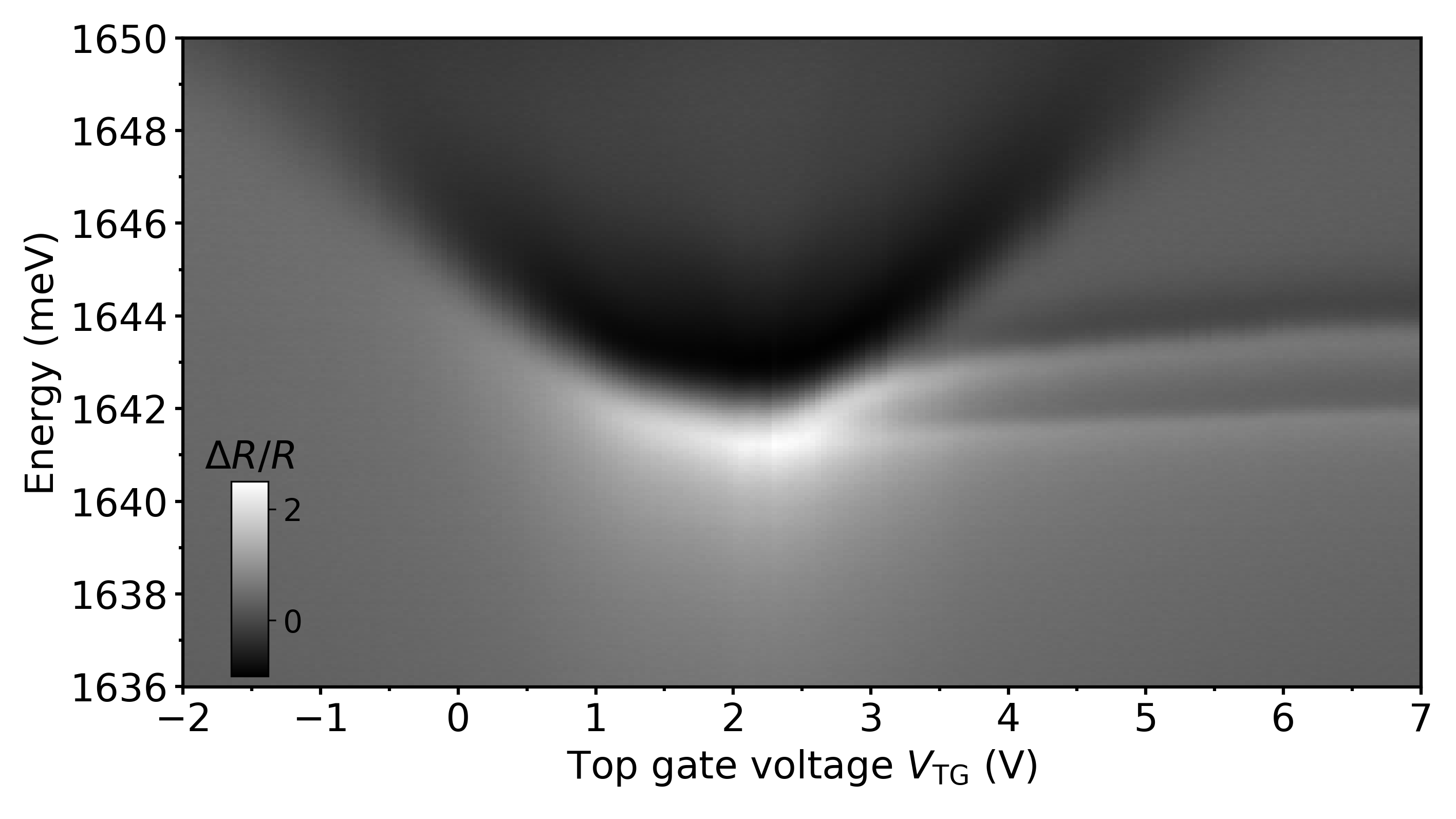}
	\caption{\textbf{Quantum confinement in the n-i-p regime.} Normalized reflectance $\Delta R/R$ from Device 2 as a function of $\vtg$ for fixed $\vbg = -7\,$V, corresponding to hole-doping in region I. We observe qualitatively similar signatures of quantum confinement as the p-i-n regime shown in the main text, which includes narrow discrete states emerging out of the repulsive polaron (RP$^-$) continuum for $\vtg \gtrsim 2.5\,$V.}  
	\label{fig:WL_n-i-p}
\end{figure}

\subsubsection*{n-i-p regime}
\noindent In our work, we mainly focused on the p-i-n regime, where region I is n-doped and region II is p-doped. In principle, the mechanism of quantum confinement in the i-region should work the same way even in the opposite n-i-p regime where region I is p-doped and region II is n-doped. In Fig.\,\ref{fig:WL_n-i-p}, we show the normalized reflectance measured as a function of $\vtg$ for fixed $\vbg = -7\,$V. 
% This measurement is performed at a position on the sample where the TG is 200 nm wide, similar to Fig.\,4\,\bfB. 
We observe similar qualitative signatures of quantum confinement, i.e.\,the narrow discrete lines emerging out of the repulsive polaron continuum. As expected, these lines now appear at positive $\vtg$. However, we do observe conspicuous quantitative differences between the p-i-n and n-i-p settings. We find that a prolonged neutral region, that is seen in Fig.\,2\,{\bfB}, is absent in the data shown in Fig.\,\ref{fig:WL_n-i-p}. Because of this we suspect that the discrete resonances in the n-i-p scenario do not exhibit the sharp initial red shift as observed in the voltage range $-6\,\mathrm{V} < \vtg < -4\,\mathrm{V}$ in Fig.\,2\,{\bfB}. 
% We currently do not understand the origin of this discrepancy.
We suspect that the origin of this discrepancy could be rooted in the substantially different leakage current for electrons and holes. Due to the small valence band offset between h-BN and MoSe$_2$ the tunnelling rate for holes is expected to be considerably larger than for electrons. This gives rise to a long lifetime of trapped electrons. Hence, while increasing $\vtg$ a non-equilibrium electron population in region II can be sustained immediately after the hole population has been fully depleted.

\subsection{Line shape analysis of reflectance and photoluminescence data}

\begin{figure*}[ht!]
	\includegraphics[width=13cm]{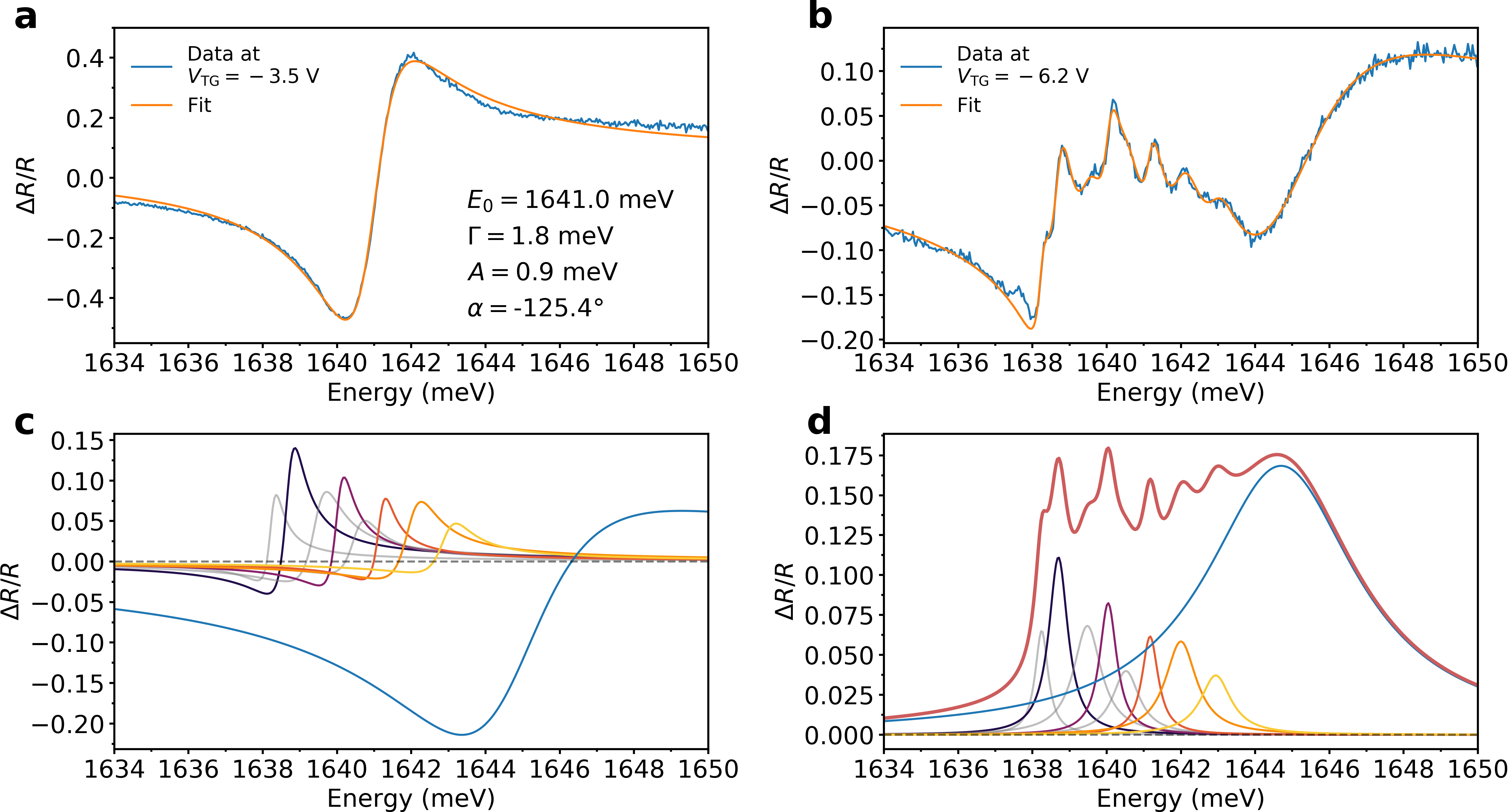}
	\caption{\textbf{Line shape analysis of reflectance data.} (\bfA) Spectral profile described by Eq.\,(\ref{eqn:spec_func}) fit to the bare 2D exciton transition at $\vtg = -3.5$\,V. (\bfB) Reflectance linecut at $\vtg = -6.2$\,V fit by a superposition of multiple narrow spectral profiles	$\sum_i S(E;E_{0,i},\Gamma_i,A_i,\alpha_c) + S(E;E_\mathrm{RP},\Gamma_\mathrm{RP},A_\mathrm{RP},\alpha_\mathrm{RP})$. (\bfC) Individual components of the fit. (\bfD) Individual components after removing line asymmetry by setting $\alpha_c$ and $\alpha_\mathrm{RP}$ to $0^\circ$. The resulting overall lineshape is shown in red. }  
	\label{fig:fits}
\end{figure*}

While a detailed analysis of the reflectance line shape requires the use of the transfer matrix method, a more straightforward approach is to describe the measured reflectance signal as $\mathrm{Im}\left[ e^{i\alpha(E)} \chi(E)\right]$, where $\chi(E)$ is the MoSe$_2$ monolayer optical susceptibility and $\alpha(E)$ a wavelength-dependent effective phase shift \cite{Smolenski2018}. The parameter $\alpha(E)$  captures the effect of light interfering at different material interfaces in our device heterostructure (e.g.\,h-BN/Au). To first order we assume $\alpha$ to be wavelength-independent in our spectral range of interest. The reflectance spectral profile $S(E)$ associated with an optical resonance can then be modelled in the following manner:
\begin{align}
% L_0(E) &= \frac{1}{\pi} \frac{\Gamma/2}{(E-E_0)^2 + \Gamma^2/4} \\ 
L_0(E) &= \frac{\Gamma/2}{(E-E_0)^2 + \Gamma^2/4} \\ 
L_D(E) &= \frac{E_0-E}{(E-E_0)^2 + \Gamma^2/4} \\
S(E) &= A \left( \mathrm{cos}(\alpha) L_0(E) + \mathrm{sin}(\alpha) L_D(E) \right) + C \label{eqn:spec_func}
\end{align}
where $L_0(E)$ and $L_D(E)$ constitute a pure Lorentzian and a dispersive Lorentzian line shape, respectively, with $E_0$ being the center frequency and $\Gamma$ the linewidth. The parameter $A$ characterizes the overall amplitude of the resonance, while $C$ takes into account any broad background signal. The result of fitting this spectral profile to the bare 2D exciton transition (corresponding to the line cut taken at $\vtg = -3.5$\,V in Fig.\,2\,\bfC) is depicted in Fig.\,\ref{fig:fits}\,\bfA.

% % \begin{figure}
% % 	\includegraphics[width=8cm]{Figures/FigS6.png}
% % 	\caption{\textbf{Comparison between PL data and prediction from electrostatic simulations.} (\bfA) Evolution of zero-point energy and (\bfB) level splitting $\hbar\omega$ between the lowest two states as a function of $\vtg$. Data obtained from simulations is depicted in blue, while experimental data points from the PL measurement are shown in red.}  
% % 	\label{fig:sim_PL}
% % \end{figure} 

We attribute the origin of the discrete spectral features in our optical measurements to quantum confinement of excitons. As a consequence, when the energy splitting $\hbar\omega$ becomes comparable to the exciton linewidth, we expect to see a coherent superposition of overlapping discrete lines which are associated with individual states splitting off from a broad continuum resonance. To justify this claim we take the spectrum at $\vtg = -6.2$\,V (Fig.\,2\,\bfC) as an example and fit it with a superposition of multiple narrow spectral profiles $\sum_i S(E;E_{0,i},\Gamma_i,A_i,\alpha_c)$, which characterize the lines associated with quantized motional states, and a broad resonance $S(E;E_\mathrm{RP},\Gamma_\mathrm{RP},A_\mathrm{RP},\alpha_\mathrm{RP})$, that accounts for the repulsive polaron continuum. As shown in Fig.\,\ref{fig:fits}\,\bfB, a good fit of the measured data can be achieved over the whole spectral range of interest. We emphasize that during this procedure the same phase factor $\alpha_c$ is assumed for all lines associated with confined states. This can also be seen in Fig.\,\ref{fig:fits}\,\bfC, which depicts the individual components of the overall spectrum. The states, for which the transition energy is not traced in Fig.\,3\,\bfD, are thereby marked in gray. Furthermore, the phase factor of the hole repulsive polaron resonance $\alpha_\mathrm{RP}$ is not the same as $\alpha_c$. This is justified considering that the origin of this resonance is rooted in a separate spatial region of the device, thus causing a different interference pattern. For the purpose of illustration, we also show in Fig.\,\ref{fig:fits}\,{\bfD} the resonances when the asymmetry in their line shape is removed by setting $\alpha_c = 0^\circ$ while retaining the other fit parameters.

The full list of parameters obtained from the fit of this spectrum at $\vtg = -6.2\,$V are shown in table \ref{table:fit_params}. The lower energy resonances exhibit a linewidth $\Gamma_i \gtrsim 500\,\mu$eV, which is almost a factor of $4$ smaller than the bare 2D exciton linewidth (see Fig.\,\ref{fig:fits}\,\bfA). At lower $\vtg$, where a stronger excitonic confinement is expected, we observe narrowing of the linewidth down to $300\,\mu$eV. This analysis shows how after the onset of hole doping, not only does the continuum blue shift due to the emergence of the repulsive polaron, but concurrently a broadening and loss of oscillator strength for the higher-lying discrete states is observed. 

Due to the intricate structure of our reflectance data, the fitting procedure described above cannot be performed in an automated way for the entire range of $\vtg$. Hence, the transition energies shown in Fig.\,2\,{\bfD} are obtained by selecting a narrow spectral range which only contains the confined states and fitting a superposition of pure Lorentzians i.e.\,$\sum_i L_0(E;E_{0,i},\Gamma_i,A_i)$ i.e.\,the phase factor $\alpha_c$ is set to $0^\circ$. This assumption likely introduces a systematic offset in the estimation of the resonance energy. However, this is not a big hindrance since we focus on the evolution of transition energies as a function of $\vtg$. Similarly, photoluminescence spectra are also fit by a superposition of pure Lorentzians i.e.\,$\sum_i L_0(E;E_{0,i},\Gamma_i,A_i)$.

% In contrast to the fitting procedure described above, the transitions energies shown in Fig.\,2\,\bfD are obtained from the reflectance spectra by 
% reflectance measurements, spectra acquired in photoluminescence measurements are fit by a superposition of pure Lorentzians i.e.\,$\sum_i L_0(E;E_{0,i},\Gamma_i,A_i)$.

\begin{table}[h!]
\centering
\begin{tabular}{||c|c|c|c|c||} 
 \hline
 i & $E_{0}$ (meV) & $\Gamma$ (meV) & $A$ (meV) & $\alpha$ ($^\circ$) \\ [0.5ex] 
 \hline\hline
 1  & 1638.2 & 0.42 & 0.022 & -55.5 \\ 
 2  & 1638.7 & 0.62 & 0.056 & -55.5 \\
 3  & 1639.5 & 0.91 & 0.050 & -55.5 \\
 4  & 1640.1 & 0.56 & 0.038 & -55.5 \\
 5  & 1640.5 & 0.88 & 0.028 & -55.5 \\
 6  & 1641.2 & 0.53 & 0.027 & -55.5 \\
 7  & 1642.0 & 1.02 & 0.048 & -55.5 \\
 8  & 1642.9 & 1.00 & 0.030 & -55.5 \\
 RP & 1644.7 & 4.95 & 0.685 & 236.8 \\ [1ex] 
 \hline
\end{tabular}
\caption{Fit parameters obtained by fitting $\Delta R/R$ at $\vtg = -6.2$\,V (see Fig.\,\ref{fig:fits}\,\bfB) }
\label{table:fit_params}
\end{table}

\subsection{Polarization anisotropy}

\begin{figure*}
	\includegraphics[width=12cm]{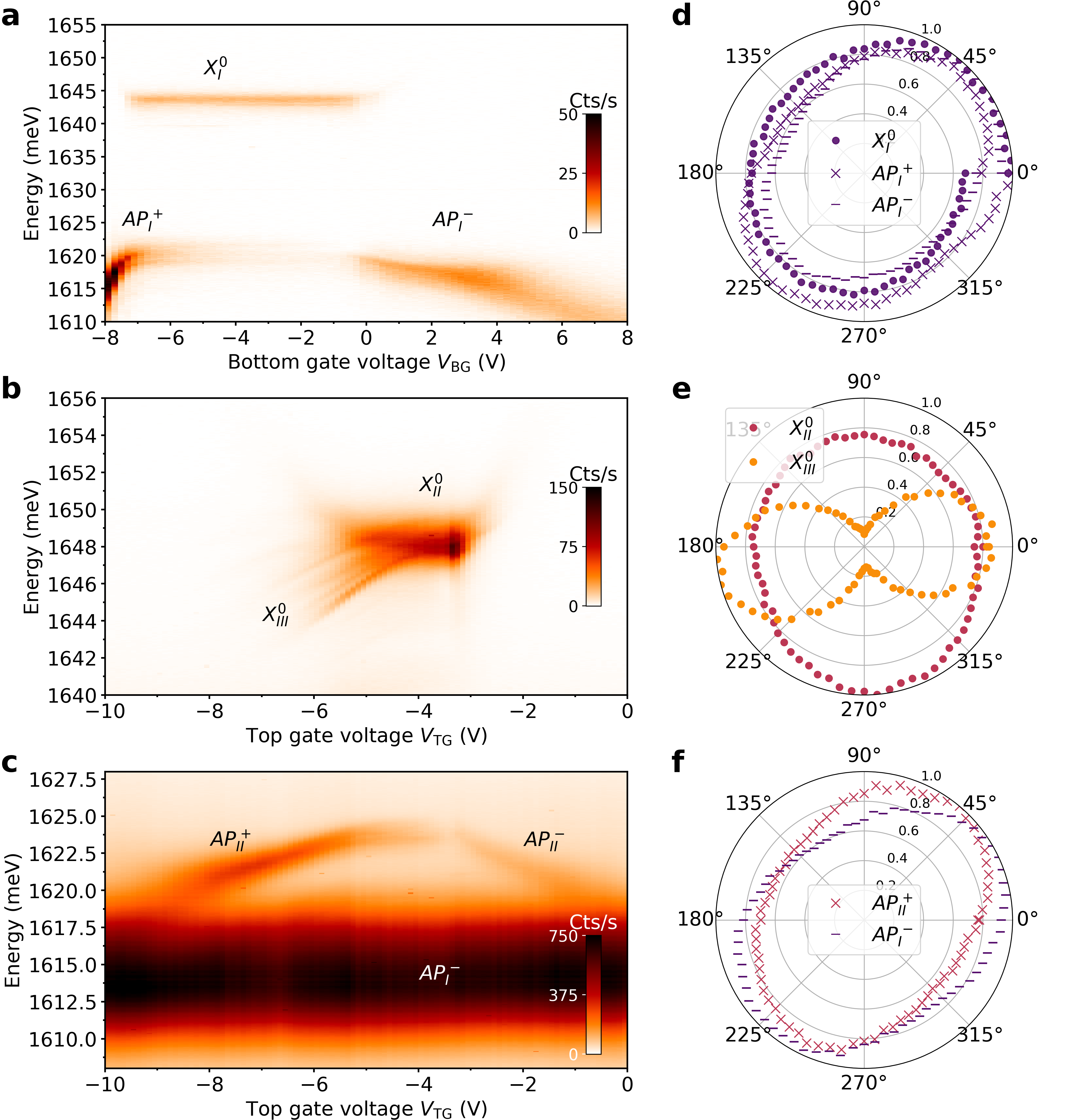}
	\caption{\textbf{Linear polarization anisotropy of different optical resonances.} (\bfA) PL bottom gate ($\vbg$) scan conducted on bare MoSe$_2$ away from top gate region. (\bfB), (\bfC) PL top gate ($\vtg$) scan performed on the split gate away from the gap region. (\bfD)-(\bfF) Polarization dependence of optical transitions in regions I, II and III, which are represented in purple, magenta and orange, respectively. The exciton, hole-side attractive polaron and electron-side attractive polaron is marked with a circle, cross and dash, respectively.}  
	\label{fig:polarization}
\end{figure*}

To verify that the strong polarization anisotropy we report for Device 1 in the main text is indeed associated with the confined states of the exciton, we perform polarization-resolved PL measurements also for various other optical transitions observed in the device. Fig.\,\ref{fig:polarization}\,{\bfA} demonstrates a reference bottom gate ($\vbg$) scan conducted at a position away from the TG region. Also shown in Fig.\,\ref{fig:polarization}\,{\bfB} and {\bfC} is a $\vtg$ scan performed on the split gate away from the gap region. We reiterate that since the spot size of our excitation beam is larger than the split gate width ($200$\,nm), PL emission from three distinct spatial regions is measured:
\renewcommand{\labelenumi}{(\Roman{enumi})}
\begin{enumerate}
	\item The region away from the TG is electron-doped and thus gives rise to a broad attractive polaron resonance ($\mathrm{AP}^-_\mathrm{I}$) centered at $\sim1.615$ eV, which remains unaffected as a function of $\vtg$.
	\item The region underneath the TG leads to neutral exciton ($\mathrm{X}^0_\mathrm{II}$) and attractive polaron resonances ($\mathrm{AP}^-_\mathrm{II}$ and $\mathrm{AP}^+_\mathrm{II}$), as it changes from being electron-doped, neutral and finally hole-doped when $\vtg$ is lowered.
	\item The neutral intermediate region at the edge of TG gives rise to red-shifting confined exciton lines ($\mathrm{X}^0_\mathrm{III}$), as $\vtg$ is lowered.
\end{enumerate} 

In Fig.\,\ref{fig:polarization}\,{\bfD}-{\bfF}, we show the polarization dependence of these optical transitions by plotting the normalized PL emission as a function of linear polarization detection angle. $0^{\circ}$ thereby constitutes the direction along the quantum wire and thus the TG edge. As shown in Fig.\,\ref{fig:polarization}\,\bfF, the attractive polaron resonances originating from underneath the TG ($\mathrm{AP}^+_\mathrm{II}$) and away from the TG ($\mathrm{AP}^-_\mathrm{I}$) have a low degree of linear polarization $\xi = (I_{max} - I_{min})/(I_{max} + I_{min}) \lesssim 20\%$, where $I_{max}$ and $I_{min}$ are the maximum and minimum intensities, respectively. These are obtained by fitting the function $A\cdot\mathrm{cos}^2(\theta-\theta_0)+C$ to the normalized PL intensities as function of the detection angle $\theta$. The neutral exciton originating from underneath the TG ($\mathrm{X}^0_\mathrm{II}$) exhibits a similar behavior ($\xi \approx 10\%$, see Fig.\,\ref{fig:polarization}\,\bfE). Furthermore, the primary polarization axis of these resonances, given by $\theta_0$, has varying orientation and does not align with the TG edge. As a reference, we also compute $\xi$ for these optical transitions away from the TG and find that it is in the same range (Fig.\,\ref{fig:polarization}\,\bfD). In stark contrast to this behavior, the confined exciton states exhibit $\xi=96\%$ along with a primary polarization axis oriented within $\pm 5^{\circ}$ along the TG edge.

Due to the similarity in polarization properties of exciton and polaron resonances from region I and II, we conclude that the strongly polarized emission of the confined excitonic states has its origin in the 1D confinement rather than a screening by the metallic TG, the effect of which cannot be distinguished from a typical strain-induced polarization dependence. If the geometry of the metallic TG had an impact on the polarization properties of the emission, also other optical transitions would exhibit an enhanced linear emission.

% % \newpage
% % \begin{figure*}
% % 	\includegraphics[width=12cm]{Figures/FigS9.png}
% % 	\caption{\textbf{Discrepancy between reflectance and PL.} In the data shown in Fig.\,3\,\textbf{D} of the main text, we observe a pronounced discrepancy between the voltage dependence of reflectance and PL lines. Here, we show the difference between the energy of the lowest state from reflectance and PL measurements (blue circles) as a function of $\vtg$ along with the hole-side Fermi energy (red diamonds). They show reasonable agreement, which suggests that the discrepancy stems from a hole-density dependent effect. We speculate that the origin could lie in the interaction between the dipolar excitons in the i-region and the charges in the surrounding regions, which manifests in different signatures in reflectance and PL.} 
% % 	\label{fig:Ef}
% % \end{figure*}

% \section{To DO:}

% \begin{itemize}
%     \item Add section on Device B; details of device, voltage parameters etc
%     \item Add section on differences between Device A and B
%     \item Improve section on confining potentials (new sims?); 
%     \item Add section on COM and relative wavefunctions, diamagnetic shifts and splitting
%     \item Improve doping properties section. 
%     \item Move section on discrepancy between reflection and PL to Supplementary and connect it to the figure of energy vs EF. 
    
% \end{itemize}

\end{document}